\documentclass[aps,prx,twocolumn,a4paper]{revtex4-2}

\usepackage{hyperref}
\usepackage{geometry}                		
\usepackage{graphicx}				
\usepackage{amsmath}
\usepackage{amssymb}
\usepackage{color}

\usepackage{ascii}
\usepackage{tikz}
\usetikzlibrary{quantikz2}
\usepackage{adjustbox}

\begin{document}
\title{Stochastic Neural Networks for Quantum Devices}
\author{Bodo Rosenhahn$^1$, Tobias J.\ Osborne$^2$, Christoph  Hirche$^1$}
\affiliation{1) Institute for Information Processing (tnt/L3S), Leibniz Universit\"at Hannover, Germany}
\author{ }
\affiliation{2) Institute for Theoretical Physics (ITP/L3S), Leibniz Universit\"at Hannover, Germany, \\
Corresponding Authors: \{rosenhahn,hirche\}@tnt.uni-hannover.de}

\date{\today}

\begin{abstract}
This work presents a formulation to express and optimize stochastic neural networks as quantum circuits in gate-based quantum computing. Motivated by a classical perceptron, stochastic artificial neurons are introduced and combined into a quantum neural network. 
The Kiefer-Wolfowitz algorithm in combination with simulated annealing is used for training the network weights. Several topologies and models are presented, including shallow fully connected networks, Hopfield Networks, Restricted Boltzmann Machines, Autoencoders and convolutional neural networks. We also demonstrate the combination of our optimized neural networks as an oracle for the Grover algorithm to realize a quantum generative AI model.  
\end{abstract}

\maketitle

\section{Introduction}

Artificial neural networks are a common and increasingly important family of mathematical models consisting of connected artificial neurons arranged in a large graphical model. They have shown impressive performance in a large series of applications, ranging from classification and regression \cite{bishop2023learning}, to data compression \cite{autoenc2018,kingma2022autoencodingvariationalbayes}, anomaly detection \cite{RudolphDifferNet} or content generation \cite{VQVAE19}. They are commonly used for supervised, unsupervised \cite{6472238}, self-supervised \cite{8237488}, semi-supervised \cite{Abbasnejad2017CVPR} or reinforcement learning \cite{vanOtterlo2012}.  Modern architectures, e.g. based on transformers \cite{Transformer17} are used in Large Language Models \cite{LLM20,LLM22}, Vision Language Models \cite{VLM24} and provide an impressive amount of information in foundation models \cite{Hollmann2025,ma2024tabdpt,FModelsGeo,Kirillov_2023_ICCV}.
\begin{figure}
  \includegraphics[width=0.45\textwidth]{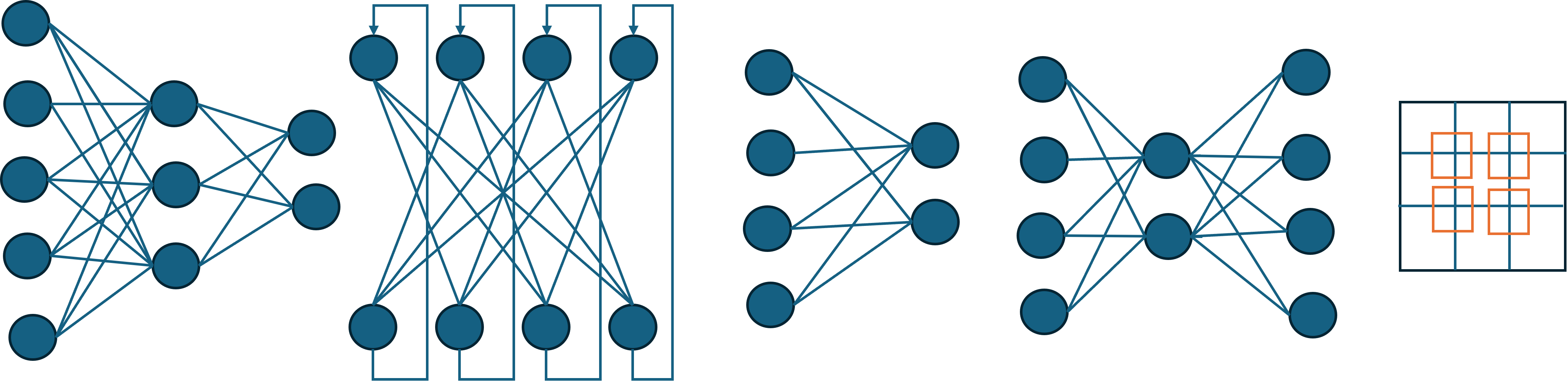}    
\caption{Realized network architectures for a quantum computer (from left to right): Shallow (fully connected) neural networks for classification, Hopfield networks for pattern memorization, Restricted Boltzmann Machines, autoencoders for representation learning and convolutional neural networks. 
}
\label{fig:NNModels}
\end{figure} 


In this work, we model probabilistic artificial neurons on a quantum device and assume a quantum state as input. 
Since a core property of a probabilistic artificial neuron is its probabilistic activation (in contrast to a deterministic activation based on an activation function), it is a well-suited model for a quantum computer. In the following sections, we describe how to model a probabilistic artificial neuron on a quantum device in a minimalistic fashion and show in several examples that our approach is flexible across different neural network topologies. We present experiments with shallow fully connected networks, Hopfield networks, Restricted Boltzmann Machines, autoencoders, and convolutional neural networks. Finally, we present an approach to combine our network with the Grover algorithm to achieve generative AI on a quantum device.
For optimization we propose to make use of simulated annealing, combined with the Kiefer-Wolfowitz algorithm, as detailed later.
Several quantum perceptron models have been proposed in the past, e.g. suitable for spiking neural network models \cite{Kristensen2021} or recurrent networks \cite{Bausch20}. However, most models either require ancilla qubits or propose a repeat-until-success procedure to mimic the deterministic behavior of an activation function \cite{cao2017quantumneuron} with a reasonably high gate count.  In contrast, we propose a simple model with a probabilistic activation property, well suited for solving different machine learning tasks. 
 Our core contributions can be summarized as follows:
\begin{enumerate}
\item  We propose a perceptron model for a quantum device. It provides a probabilistic activation, is free of any ancillary qubits and the proposed artificial neurons can be arranged in a layered structure, similar to a classical neural network.
\item We present how to model and optimize neural networks for a quantum device, based on simulated annealing and the Kiefer-Wolfowitz algorithm. The approach is well suited to ensure additional constraints such as shared weights or specific connections which is useful for optimizing different architectures.
\item Several experiments on different architectures, including shallow neural networks, Hopfield Networks, Restricted Boltzmann Machines, autoencoders, and convolutional neural networks demonstrate the general applicability of the proposed model. Our experiments are  based on binary input patterns, but the framework can be easily extended to more general qubit states, as clarified later.
\item We demonstrate the combination of our circuit architecture as an oracle for a Grover algorithm to make use of a trained (frozen) network for GenAI. 
\end{enumerate} 

There are two core arguments why it is interesting to realize machine learning models directly on a quantum computer: Firstly, in combination with quantum sensing, the sensed information is already in the quantum domain (e.g. encoded in the qubit states) and it is useful to stay in this domain for further postprocessing. A standard framework requires readouts to map the information to the digital domain and from thereon, signal analysis or machine learning inference can be performed~\cite{SinananSingh2024singleshotquantum,liu2025mixed}. It is well known that the process of digitization, or if needed full tomography, can be tedious and might require a large amount of readouts before the data can be processed reliably~\cite{Giarmatzi2025multitimequantum}. Instead, our model can directly pick up the quantum signal and continue to process (e.g. to classify) it. Thus, our work provides a framework on how to handle and process sensed quantum information on-board.
A second important aspect is that even though our model itself does not provide a direct quantum advantage, in combination with known approaches indicating a quantum advantage it can provide powerful new methods:
Many approaches in GenAI (e.g. diffusion models \cite{rombach2022high}) require the repetitive evaluation of a neural network, e.g. to denoise an image. In contrast, our proposed neural network can be used as an oracle for the Grover algorithm. This allows us to generate examples just by evaluating the circuit once.
Note that some approaches from GenAI, e.g. based on adversarial networks \cite{Sengar2025} suffer during training from a so-called mode collapse. This does not happen in our quantum generative model.
We clarify here, that the existing state of the art neural network models are huge, realized with billions of parameters. As the amount of available qubits is limited, and the qubits suffer from decoherence, models at a similar scale will only become practically available on possible future fault-tolerant large scale quantum devices.

\section{Fundamentals}
\subsection{The Perceptron}
\label{SecFPerc}
In 1943, McCulloch and Pitts formulated their idea for logical calculus using concepts from nervous activities~\cite{McCulloch1943}. 
For a McCulloch-Pitts cell with $n$  exciting input lines on which the signals $(x_1 \ldots x_n)$  are applied, and $m$  inhibiting input lines on which the signals $(y_1 \ldots  y_m)$  are applied, the calculation works as follows: If $m \geq 1$ and if one of the signals $(y_1 \ldots y_m)$ equals 1, the artificial neuron outputs a $0$. 
Otherwise, the input signals 
$(x_1 \ldots x_n)$ are summed up to  $x=\sum x_i$. For $n = 0$, 
$x = 0$  is set. The value $x$  is compared to the threshold $\theta$. If the value $x$ is greater than or equal to $\theta$, the artificial neuron returns $1$, otherwise it returns $0$.
In 1958, Frank Rosenblatt published his perceptron model which extends the summation to a scalar product, followed by a step function~\cite{rosenblatt1958}.  
The perceptron can be summarized as
\begin{equation}
\sigma_j = \left\{ \begin{array}{l}
1 : \sum_i \omega_{ij}x_i+b > 0\\
0 : else
\end{array}
\right.
\end{equation}
The bias value $b$ corresponds to the decision threshold and $\omega_{ij}$ are learnable parameters. 
A combination of such perceptrons in a directed acyclic graph leads to a classical (e.g.\ fully connected) neural network.
Please note, that the term artificial neuron (often abbreviated as neuron), or perceptron is often used synonymously in the literature. 
\subsection{Stochastic Artifical Neurons}
Stochastic artificial neurons are perceptron models that incorporate a random component into their activity, which distinguishes them from deterministic perceptrons. Instead of providing a fixed output for a given input, stochastic artificial neurons are activating based on a probability. This behavior is similar to the random fluctuations that occur in biological neurons, such as the random emission of neurotransmitters. Several works address this model and its optimization \cite{Müller1995,Turchetti04}. In \cite{Dutta2022} an extension towards a neural sampling machine (NSM) which exploits the stochasticity in the synaptic connections for approximate Bayesian inference has been proposed.
A stochastic artificial neuron can be formulated as a scalar product of the input values and (learnable) weights with an added bias. The score gives the probability for the activation of the neuron,
\begin{eqnarray*}
p(\sigma_j =1) &=& p\left( \sum_i \omega_i a_i + b \right). 
\end{eqnarray*}
In this work, we assume binary input values of qubits which means that $a_i \in \{0,1 \}$. To this end, the probabilistic artificial neuron can be seen as a modified version of the perceptron model, presented in \cite{cao2017quantumneuron}. The core differences are the probabilistic nature of activation, the absence of ancilla qubits, and no need for a repeat-until-success (RUS) circuit. Since the sine function is already nonlinear, our model shares similarities with Radial Basis Function networks \cite{Park91}. 

\subsection{Quantum Computing}
This work does not provide a detailed introduction to qubits and quantum gates. Please refer to the standard literature in this field \cite{10.5555/1206629}.
We assume that a quantum computer has $N$ \emph{logical} qubits, forming a quantum register. We further assume that the device is equipped with a universal gate set to express arbitrary quantum logic gates. Formally, our system can be modeled as a Hilbert space given as $\mathcal{H}\equiv (\mathbb{C}^2)^{\otimes N} \cong \mathbb{C}^{2^N}$.  A quantum circuit is a sequence of quantum gates which can be evaluated as a series of unitary matrix multiplications. In this work, two gates are mainly used: (a) a rotation gate and (b) a controlled rotation gate. The RX-Gate is a single-qubit rotation through angle $\theta$ around the x-axis. The CRX-gate is a two-qubit gate and it applies a controlled x-axis rotation of a target qubit based on the state of a control qubit. If the control qubit is in the $|0 \rangle$  state, then this gate does nothing. If the control qubit is in the $|1 \rangle$ state, then this gate rotates the target qubit state around the x-axis by an angle $\theta$.

\subsection{Quantum Neural Networks}
\begin{figure*}
  \includegraphics[width=0.9\textwidth]{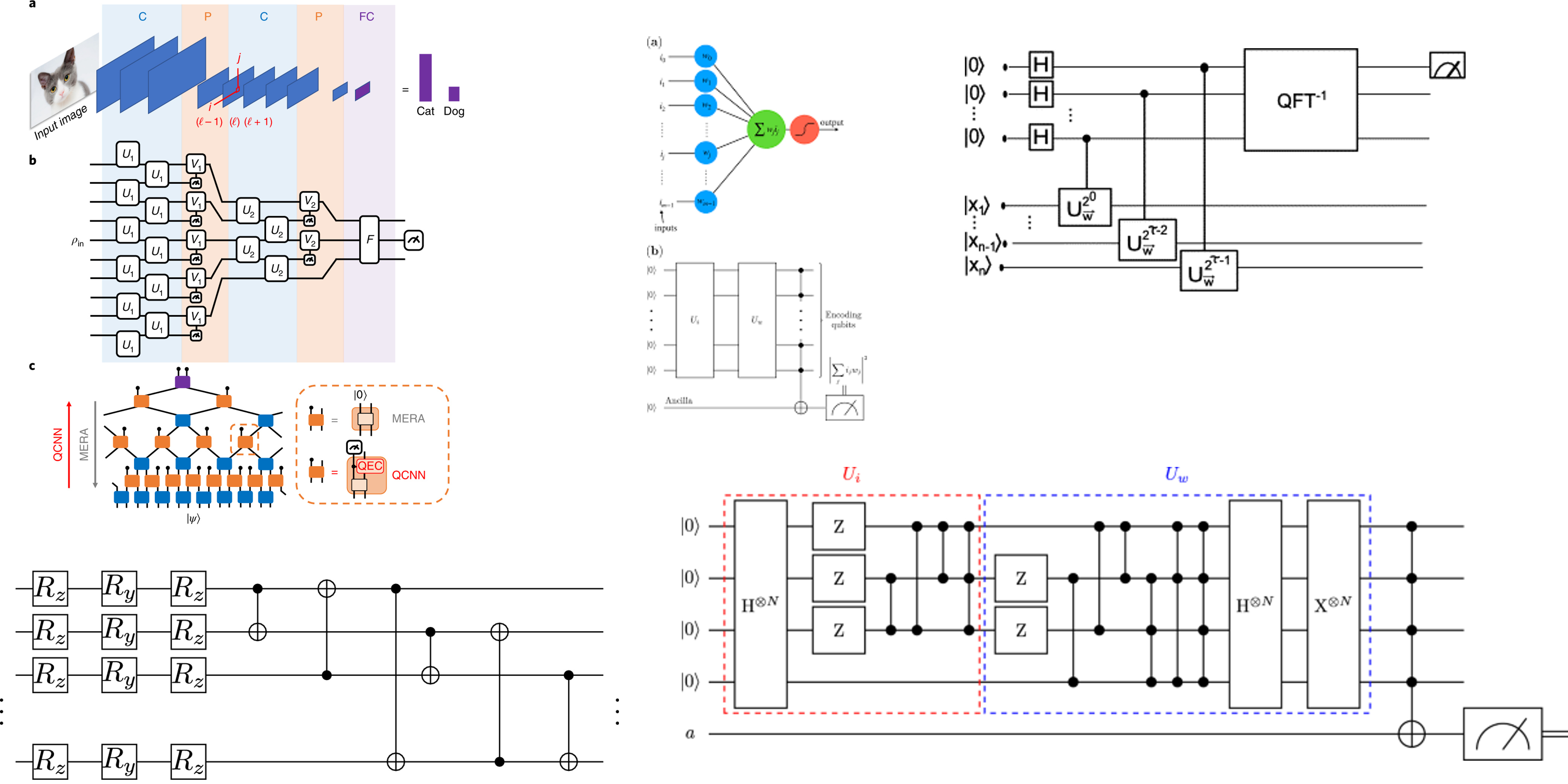}   
\caption{Examples for existing architectures to represent artificial neural networks on a quantum device (taken from \cite{Cong2019,Tacchino2019,SCHULD2015660,Bai2023}).
}
\label{fig:ExAr}
\end{figure*}

A quantum neural network (QNN) is a mathematical model that combines concepts from quantum computing and artificial neural networks. Neural networks are inspired by biological neurons which have a simple design (e.g. a scalar product followed by an activation function) which is easy to optimize (e.g. using gradient descent) but is very powerful when combined to larger networked structures, so that they can be used as generalized function approximators \cite{KAK1995259}. Several works have explored these properties by defining parameterized circuits \cite{cao2017quantumneuron} and by modeling components of typical deep models, such as convolutions \cite{bermejo2024quantumconvolutionalneuralnetworks}, recurrent neural networks \cite{Bausch20} for time-series data, or spiking neurons \cite{Kristensen2021}. The work \cite{Beer2020} defines neural networks by modeling each input value and neuron using separate qubits and by defining local unitary functions that activate the neurons while processing the input data. While these approaches have produced valuable results, they differ in important ways from classical neural network processing. For example, the series of unitary matrices used in \cite{Beer2020} makes the processing chain non-commutative, which differs from the classical approach of processing neurons independently and in parallel within a layer. Other models require ancilla qubits or repeat-until-success circuits \cite{cao2017quantumneuron}. Figure \ref{fig:ExAr} gives a collage of examples of commonly used realizations of artificial neurons and artificial neural networks.
Motivated by this, we propose an alternative model that is closer to the classical perceptron, easier to interpret and connect, and still powerful enough to map a neural network onto a quantum computer and to optimize it. We therefore represent an artificial neuron using controlled rotation gates, use layers in the neural network where perceptrons commute within each layer (as in a shallow neural network), and propose an optimization algorithm that allows us to enforce constraints such as connection cutting and weight sharing. This leads to an efficient realization of classical architectures such as Hopfield networks, restricted Boltzmann Machines, convolutions and more.
In our experiments we mainly use binary encodings and each perceptron is realized by CRX-gates. Please note, that for the case of complex input data, e.g. stemming from a quantum sensor, the CRX-gates can easily be extended to CRY- and CRZ-gates for quantum signal analysis.

\section{Quantum perceptron with probabilistic activation}
\begin{figure}
  \includegraphics[width=0.45\textwidth]{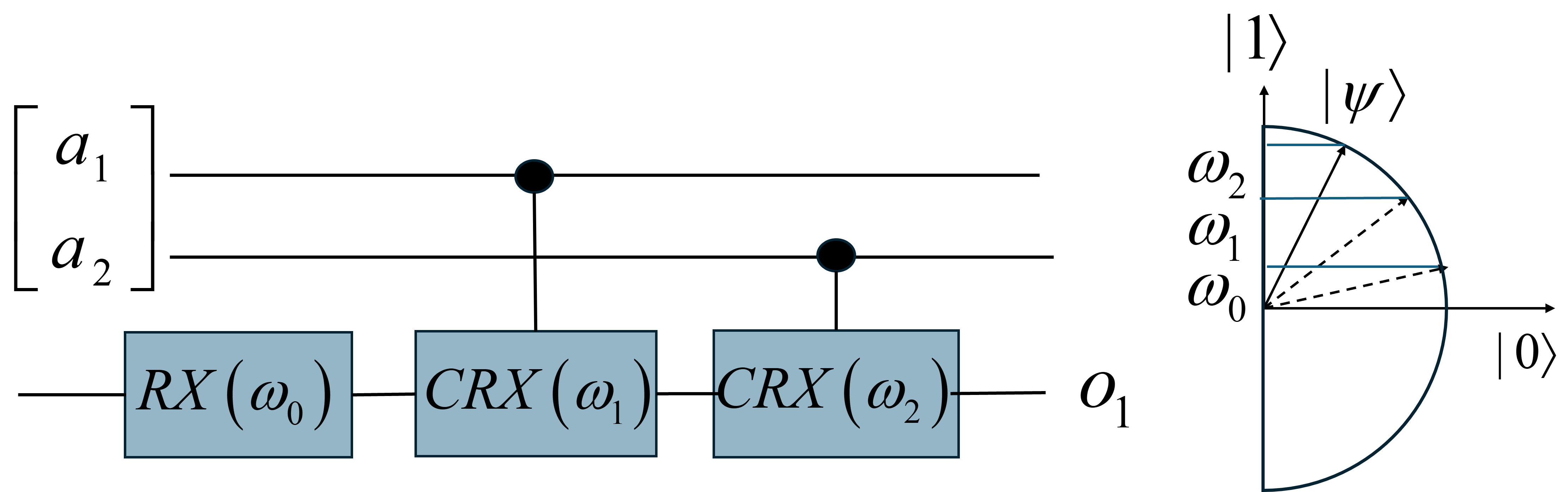}    
\caption{Concept of a quantum perceptron with probabilistic activation. The bias is represented as an RX-Gate and the binary input qubits $a_i$ increment the probability of neuron excitation by their angular components $\omega_i$ using a controlled gate. To ensure a linear and additive behavior of the angles, directly mapping to the probabilities, the $asin$-function is used. The right image visualizes the qubit activation on the Bloch sphere (just the real part is shown) and the effect of rotating the qubits, based on the bias and CRX-gates. Please note, that for complex input signals, the model can be extended with CRY- and CRZ-gates in a straightforward way.
}
\label{fig:QPerc}
\end{figure} 

\begin{figure}
  \includegraphics[width=0.45\textwidth]{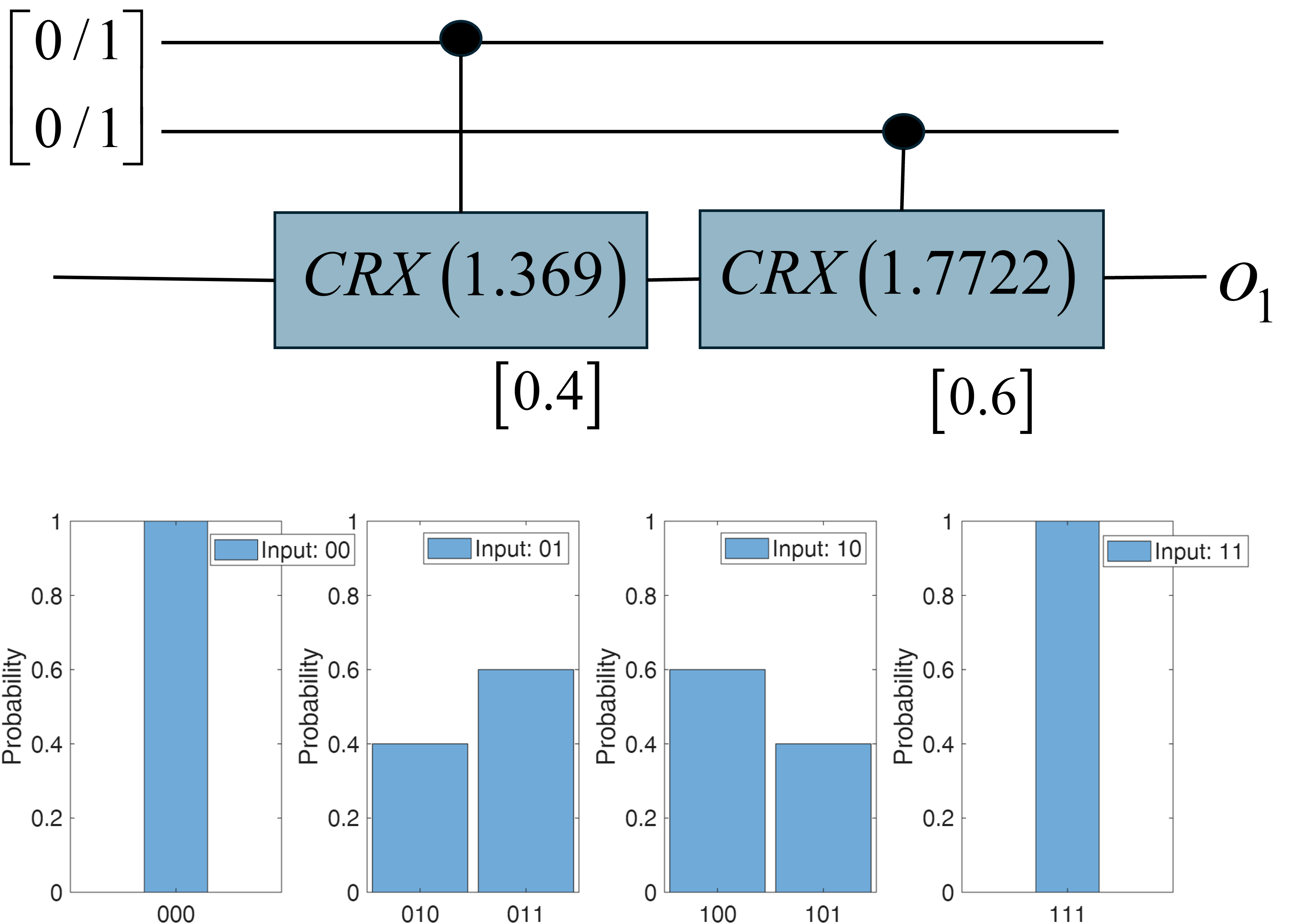}    
\caption{Example activations for a perceptron with two binary inputs. The weights are set to [0.4 0.6]. The bottom histograms demonstrate the additive increase of the activation probability. 
}
\label{fig:QPercEx}
\end{figure}

The general idea about modeling the behavior of a perceptron with probabilistic activation on a quantum computer is shown in Figure \ref{fig:QPerc}: The (binary) inputs are manipulating the activation probability of the perceptron by using RX-Gates (for the bias) and Controlled RX-Gates. The aim is to use input qubits to increase the probability of the perceptron activation by a value $a_i$. To translate this probability score to an appropriate rotation we simply use $asin(\sqrt{a_i})$ as angular measure for the rotation. The additive impact is shown in Figure \ref{fig:QPercEx}: For the first qubit we set as activation probability the value $0.6$ and for the second one the value $0.4$. The combination of all possible states $ (00, 10, 01, 11) $ leads to the observation probabilities shown in the histograms below:
If no qubit is activated, the probability to measure $0$ is one, the latter qubit raises the score to $0.6$ and if only the front qubit (the upper one) is activated, the probability of the perceptron to be activated is $0.4$. Finally, if both qubits are $1$, the perceptron activates with probability $1.0$. 
Please note, that we mainly experiment with a binary representation of data, coming from the digital domain, which gives raise to the use of a CRX-gate. For complex input signals, e.g. stemming from a quantum sensor, the perceptron can be straightforward extended to CRY- and CRZ-gates.

Please note that cumulative values greater than $1$ lead to a decrease of the activation probability. This allows the perceptron to be more expressive than a classical perceptron with e.g. a sigmoid activation. For example, the weights
$(\omega_1=1, \omega_2=1)$ will lead to the implementation of a xor-function,
\begin{eqnarray}
    |00\rangle &\rightarrow & |0\rangle \nonumber \\ 
    |01\rangle &\rightarrow & |1\rangle \nonumber \\
    |10\rangle &\rightarrow & |1\rangle \nonumber \\
    |11\rangle &\rightarrow & |0\rangle.
    \label{eq:xor}
\end{eqnarray}
This can also be motivated from medical findings, as Gidon and colleagues \cite{Gidon20} found that there exists a type of pyramidal neuron in the human cerebral cortex that can learn the XOR
function. This is  impossible with single artificial perceptrons using sigmoidal, ReLU, leaky ReLU, PReLU, GELU, or other (typical) activation functions. As mentioned before, the sin-function is already non-linear and our model shares similarities to networks which use Radial-Basis functions for activation \cite{Park91}. Please also note that in the digital domain, dropout layers are a common approach to realize probabilistic activations \cite{pmlr-v48-gal16}, but it also requires repetitive evaluations of the same (large) model to gain a probability distribution of activations. 
 
\subsection{Optimization of Quantum Neural Network Weights}
\label{ref:Opt}
For optimization of the Quantum Neural Network we use an approach known from stochastic approximation:
The Kiefer–Wolfowitz algorithm was introduced in 1952  \cite{Kiefer52} and was motivated by the  Robbins–Monro algorithm \cite{Robbins1951}. The algorithm has been presented as a method which can stochastically estimate the maximum (or minimum) of a function. Let $M(x)$ be a function which has (wlog) a maximum at the point 
$\theta$.  It is further assumed that $M(x)$ is reasonably smooth.
The structure of the algorithm follows a gradient-like method, with the iterates being generated as
\begin{eqnarray*}
         x_{n+1}&=&x_{n}+a_{n}\left(
         \frac {M(x_{n}+c_{n})-M(x_{n}-c_{n})}{2c_{n}}\right)
\end{eqnarray*}
where $M(x_n+c_n)$ and  $M(x_n-c_n)$
are independent. Thus, the gradient of $M(x)$ is approximated by a central difference method with a damping factor $a_n$ slightly shifting the solution towards an optimum. Our method works in a simulated annealing scheme:
Simulated Annealing (SA) is a probabilistic technique to approximating the optimum of a given function \cite{Kirkpatrick1983}. The name derives from annealing in metallurgy where the process involves heating and a controlled cooling of a material to change and control its physical properties. 
As an optimization scheme the algorithm works iteratively with respect to time $t$ given a state $x_t$, similar to the Kiefer-Wolfowitz algorithm. 
At each step, the simulated annealing heuristic samples a neighboring state $\hat{x}_t$ of the current state $x_t$, e.g. based on an approximate gradient. Then a probabilistic decision is made to decide whether to move to the new state $x_{t+1}=\hat{x}_t$ or to remain in the former state $x_{t+1}=x_t$.
The probability of making the transition from the current state $x_t$ to the new state $\hat{x}_t$ 
is defined by an acceptance probability function 
$P(e(x_t), e(\hat{x}_t), T)$. The function $e(x)$ evaluates the energy of this state, which is in our case 
the fitness score given by the optimization task (e.g. the $\ell_2$-loss). The parameter $T$ is a time-dependent variable dictating the behavior of the stochastic process according to a cooling scheme or annealing schedule. The $P$ function is typically chosen in such a way that the probability of accepting an uphill move decreases with time and it decreases as the difference $e(\hat{x}_t)-e(x_t)$ increases.
Thus, in contrast to a strict gradient descent, 
a small increase in error is likely to be accepted so that local minima can be avoided over the iterations, whereas a larger error increase is not likely to be accepted. A typical function for $P$ takes the form
\begin{equation}
P(e(x_t), e(\hat{x}_t), T) \propto \exp \left(  
-\frac{e(\hat{x}_t)-e(x_t)}{k T}\right),
\end{equation}
with $k$ a \emph{damping factor} $k>0$.
Thus, our optimization is based on a local stochastic search scheme which is no strict gradient descent. 
Please note that several neural networks have special requirements on the networks, such as shared weights or symmetry properties. Using stochastic search, as proposed here, can ensure certain constraints during optimization.

We will compare this approach to standard gradient descent based optimization using auto differentiation. An alternative realization of gradient descent can be based on the parameter shift rule \cite{Wierichs2022generalparameter}.
If the gate G has exactly two distinct eigenvalues $\pm r$ (which applies to all standard Pauli rotation gates such as \(R_x, R_y, R_z\) with \(r = \frac{1}{2}\)), the exact derivation is given as
\begin{eqnarray}
\frac{\partial f(\theta)}{\partial \theta} &=& r \left[ f\left(\theta + \frac{\pi}{4r}\right)\nonumber 
- f\left(\theta - \frac{\pi}{4r}\right) \right].
\end{eqnarray}
For the controlled rotation gates, four eigenvalues exist and 
the exact derivation of a CRX gate is given as
\begin{eqnarray}
\frac{\partial f(\theta)}{\partial \theta} &=& 
\frac{\sqrt{2} + 1}{4 \sqrt{2}} \left[ f\left(\theta + \frac{\pi}{4}\right) - f\left(\theta - \frac{\pi}{4}\right) \right] \nonumber \\
&&- \frac{\sqrt{2} - 1}{4 \sqrt{2}} \left[ f\left(\theta + \frac{3\pi}{4}\right)
 - f\left(\theta - \frac{3\pi}{4}\right) \right]. \nonumber \\
\end{eqnarray}
It should be noted that the parameter shift rule requires to execute the quantum circuit 2, resp. 4 times. Thus, optimization using the parameter shift rule can become compute intensive but the parameter shift rule is appropriate  when analytic gradients are feasible and simulation/runtime is not prohibitive.
\section{Models and Experiments}

\subsection{Noise Analysis of the Quantum Perceptron}

To get a better understanding of the stability of the perceptron under gate noise, especially amplitude damping, we 
assume a simple perceptron realizing a xor-decision (see equation \ref{eq:xor}). This perceptron takes two input values of \ket{q_0} and \ket{q_1} and the output qubit \ket{q_2} is realized by two CRX-gates performing a controlled $\pi/2$-rotation.  Figure \ref{fig:xorNeuron} summarizes the circuit. 
The qubits are parametrized as
\begin{align}
    \ket{q_i}=\alpha_i\ket{0}+(1-\alpha_i)\ket{1}
\end{align}

\tikzset{
 noisy/.style={starburst,draw=red,line
    width=2pt,inner xsep=-4pt,inner ysep=-5pt}
}

\begin{figure}[htb]
\resizebox{0.45\textwidth}{!}{%
\begin{quantikz}
  \lstick{\ket{q_0}} & \ctrl{2}       &  \gate[3,style={noisy},label
    style=cyan]{\mathcal{L}_1} & \qw            & \gate[3,style={noisy},label
    style=cyan]{\mathcal{L}_2}        &          & \qw \\    
    \lstick{\ket{q_1}} & \qw            & \qw                  & \ctrl{1}       &  & & \qw \\
    \lstick{\ket{q_{2}}} & \gate{RX(\pi)} &  & \gate{RX(\pi)} &  &\meter{} & \qw
\end{quantikz}}
\caption{Setup for measuring the impact of noisy gates for the implementation of an artificial neuron realizing a xor-decision.}
\label{fig:xorNeuron}
\end{figure}
\begin{figure}
  \includegraphics[width=0.45\textwidth]{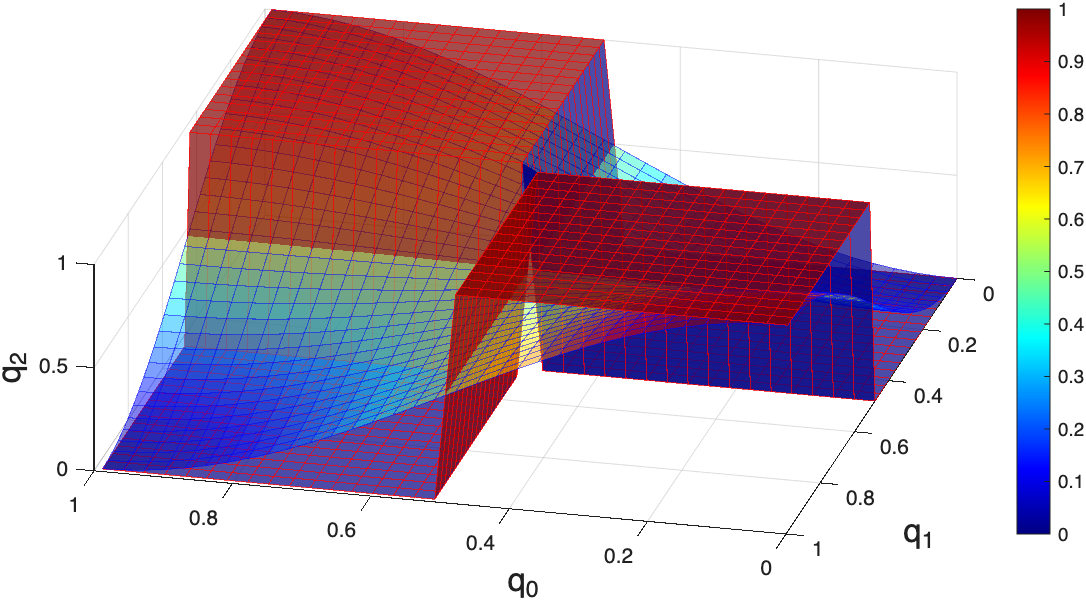}    
\caption{Probability of activating \ket{q_{2}} (Fig \ref{fig:xorNeuron}) for varying   \ket{q_{0}} and \ket{q_{1}}. Thresholding the probabilities leads to the desired xor pattern for non-binary inputs (no noise on gates).
}
\label{fig:ResxOR}
\end{figure}

Figure \ref{fig:ResxOR} visualizes the probability of activating \ket{q_{2}} for varying  non-binary values \ket{q_{0}} and \ket{q_{1}} without gate noise in the circuit. The outcome in \ket{q_{2}} is a probability of activation. Thresholding the \ket{q_{2}} measurements leads the desired xor-pattern with 

\begin{eqnarray*}
 \ket{q_0};\ket{q_1} \in [0 ... 0.5 ;0 ... 0.5  ]
    &\rightarrow & 0   \\ 
\ket{q_0};\ket{q_1} \in [0.5 ... 1 ;0 ... 0.5  ]
    &\rightarrow & 1   \\ 
 \ket{q_0};\ket{q_1}\in [0\ldots 0.5 ;0.5\ldots 1  ]
    &\rightarrow & 1  \\ 
  \ket{q_0};\ket{q_1}\in [0.5\ldots 1 ;0.5\ldots 1  ]
    &\rightarrow & 0.
\end{eqnarray*}
Thus, for non-binary values under stable gates, the circuit is performing the desired decision pattern accordingly.

\begin{figure}
  \includegraphics[width=0.45\textwidth]{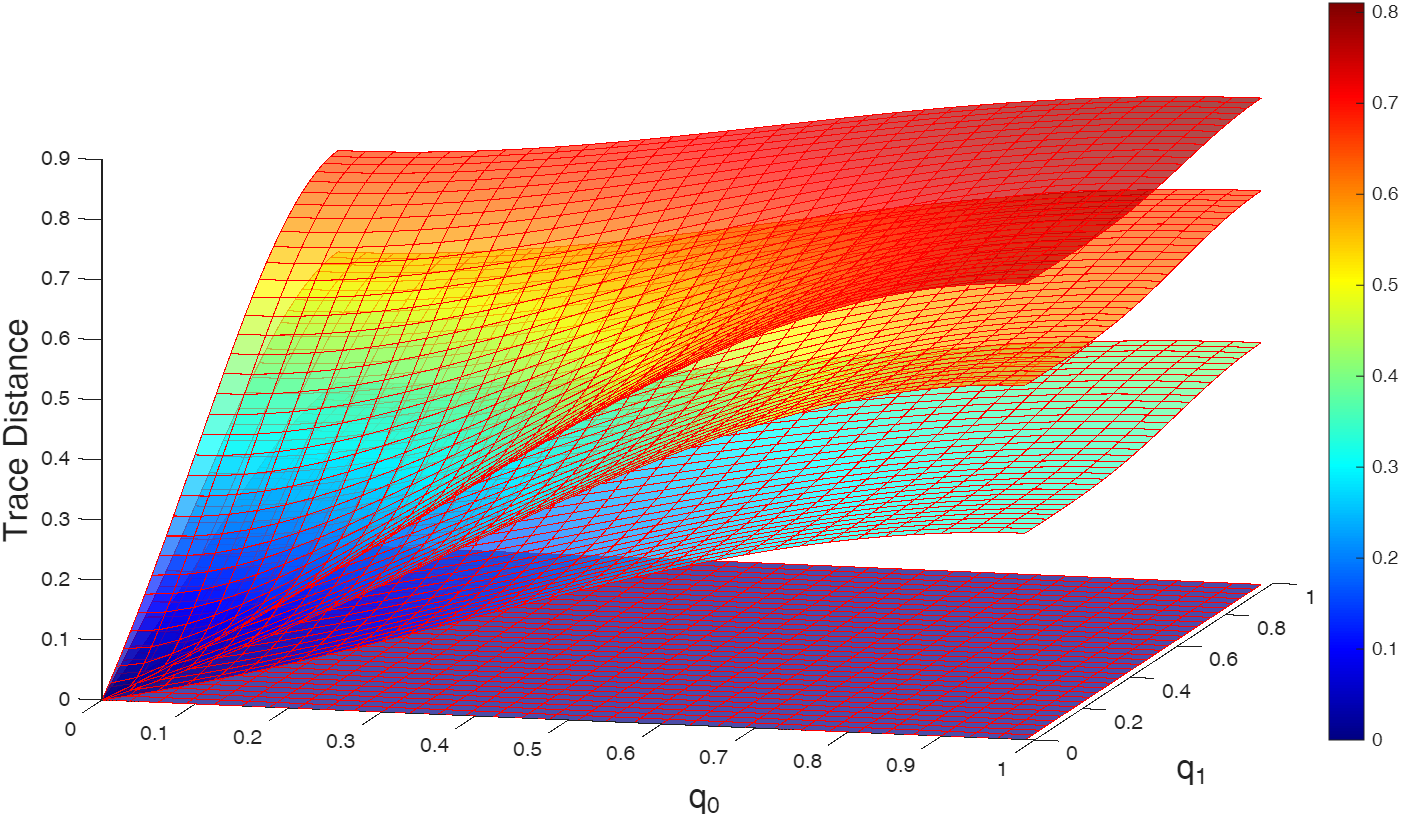}    
\caption{ Trace distances of the quantum circuit from Fig. \ref{fig:xorNeuron} when the gates are effected by noise $\mathcal{L}_1$ and $\mathcal{L}_2$.
}
\label{fig:ResGNoise}
\end{figure} 

To model energy relaxation effects in the three-qubit system, we employ an amplitude damping channel \cite{nielsen2000quantum}. The standard single-qubit amplitude damping process can be characterized by a set of Kraus operators $\{E_0, E_1\}$, defined as:
\begin{equation}
    E_0 = \begin{pmatrix} 1 & 0 \\ 0 & \sqrt{1-\gamma} \end{pmatrix}, \quad 
    E_1 = \begin{pmatrix} 0 & \sqrt{\gamma} \\ 0 & 0 \end{pmatrix},
\end{equation}
where $\gamma \in [0, 1]$ represents the damping probability (or decay rate) of a single qubit from the excited state $|1\rangle$ to the ground state $|0\rangle$. 

Now, the noise mechanism during each gate operation is simulated. 
The joint noise action is represented using $2^3$ Kraus opreators
\begin{align}
    N_{i,j,k} &= E_i \otimes E_j \otimes E_k, \hspace{0.5em} i,j,k\in\{0,1\} 
\end{align}
Under this noise model, the transformation of the three-qubit density matrix $\rho$ is given by the quantum channel:
\begin{equation}
    \mathcal{N}(\rho) = \sum_{i,j,k\in\{0,1\} } N_{i,j,k} \rho N_{i,j,k}^\dagger
\end{equation}
We assume, that this effect can happen after each CRX-gate in the circuit. Thus, noise after the first gate will accumulate to the noise of the second gate.
To measure the stability, the trace distance between the unitary of the noisy circuit and the noiseless circuit is used. 
Figure \ref{fig:ResGNoise} shows the trace distance for $\gamma \in \{ 0, 0.1, 0.2, 0.3 \} $. Since amplitude decay is irrelevant for \ket{q_0} and \ket{q_1} at zero, the trace distance is very small here, but when both qubits are close to 1, the noise has an increased impact.
Since the amount of gates scale linear with the feature dimension it is obvious that for larger systems, the noise will have a major impact on the quality of the results. Thus, error correction needs to be considered when realizing the circuits in today's available hardware.

\subsection{Shallow Neural Networks}
\begin{figure}
  \includegraphics[width=0.45\textwidth]{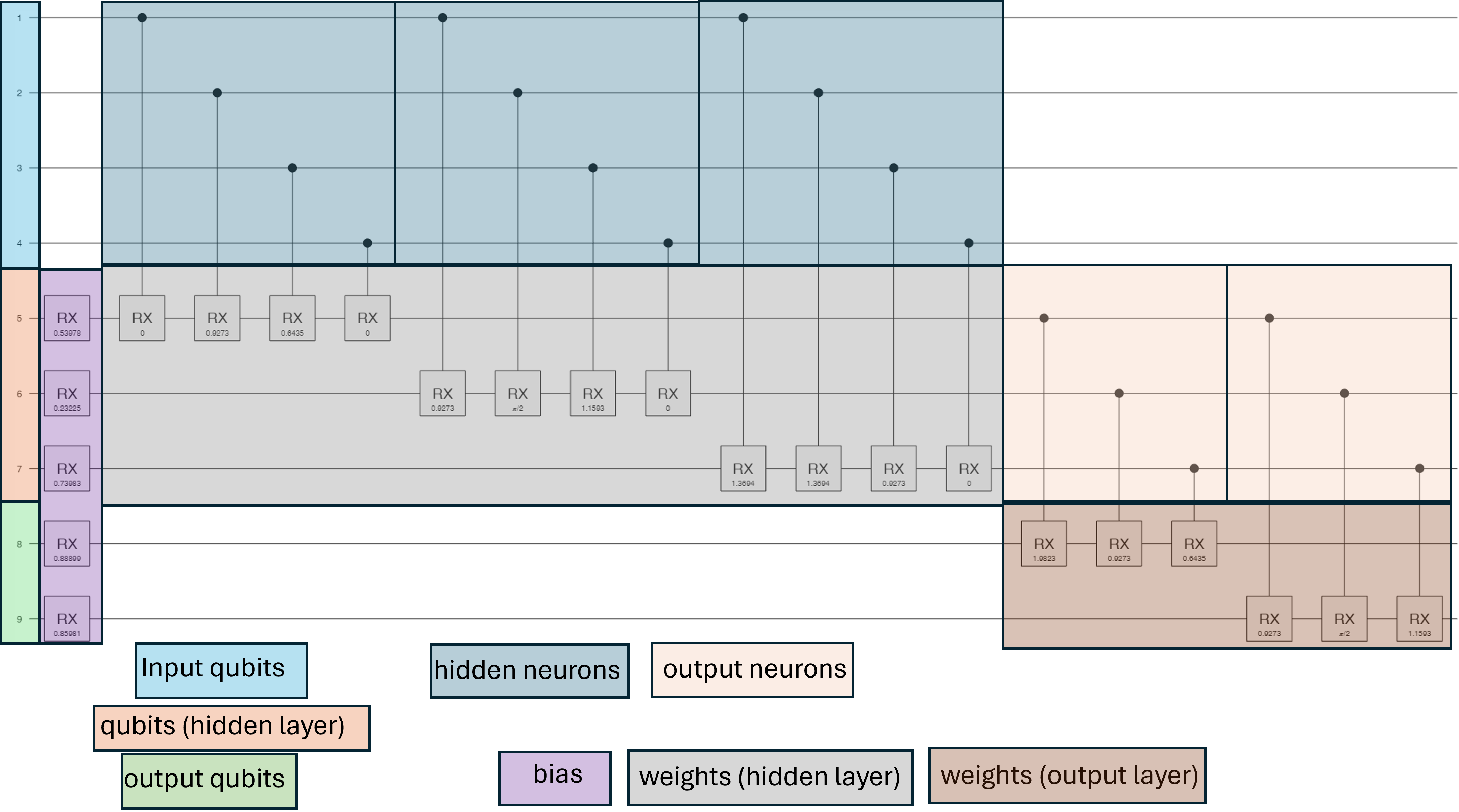}    
\caption{Implementation of a shallow neural network on a quantum device. This model consists of 4 qubits as input, three hidden artificial neurons and two output neurons. The (C)RX-Gates contain the learnable parameters. 
}
\label{fig:ShallowNN}
\end{figure}

A shallow neural network is typically a neural network that consists of only one hidden layer between the input and output layers \cite{Bishop2006,bishop2023learning}. The general structure is to connect all input elements to each perceptron of the hidden layer which is then called a fully connected layer. Afterwards the hidden perceptrons are connected to the output layer.  Figure \ref{fig:ShallowNN} demonstrates the implementation of a shallow neural network on a quantum device based on our earlier defined stochastic perceptrons. The shown model consists of 4 qubits as input which are connected to three hidden artificial neurons by using CRX-Gates. The hidden neurons are three additional qubits. The RX-Gate itself is the bias. Afterwards, the three qubits representing the neural activity are connected to two qubits acting as output neurons. The (C)RX-Gates contain with the rotation angles $\omega_i$ the learnable parameters. 

\begin{figure}
  \includegraphics[width=0.45\textwidth]{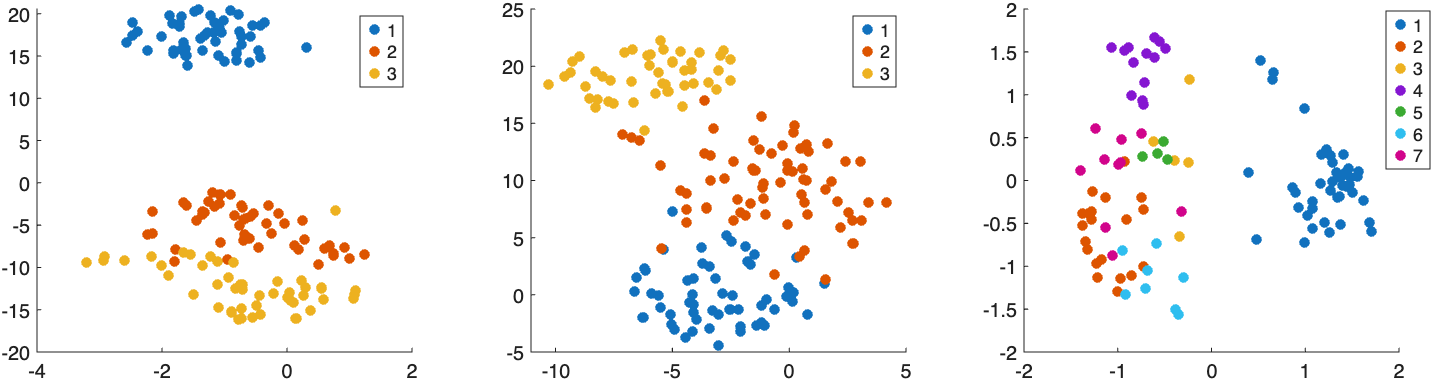}  
\caption{TSNE-Plots of the used UCI datasets for classification, the iris dataset (left), the wine dataset (middle) and the zoo dataset (right).
}
\label{fig:VisData}
\end{figure} 

\begin{figure}
  \includegraphics[width=0.45\textwidth]{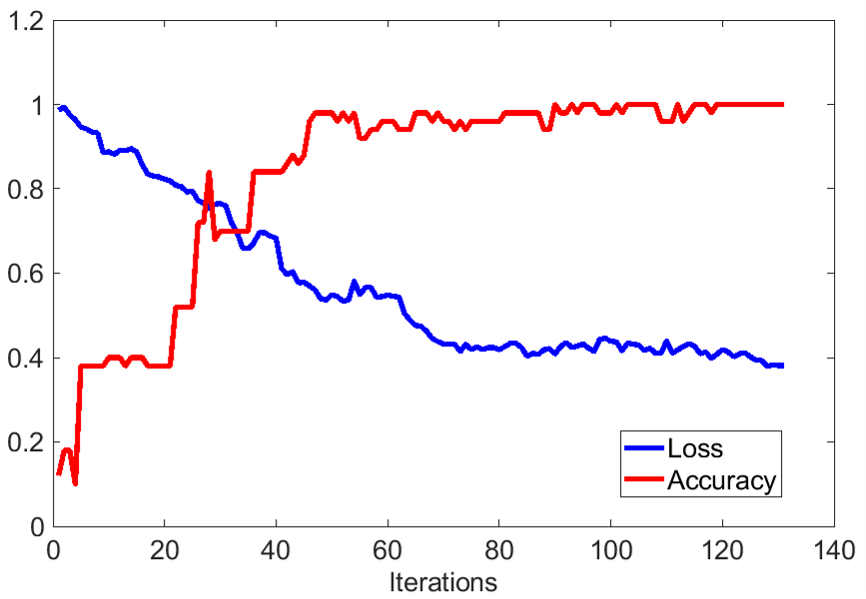}  
\caption{Convergence of the optimization algorithm during training the iris dataset. Shown is the loss decrease in blue and the accuracy (of the test data) in red. In this run, the algorithm could perfectly classify all test samples.
}
\label{fig:ConvIris}
\end{figure} 

We used the optimization procedure summarized in Section \ref{ref:Opt} and 
for the experiments the classical \textit{iris}, \textit{wine}, \textit{zoo} and  \textit{mnist} datasets were used. The datasets present multi-class classification tasks, with three categories for the iris and wine datasets and seven categories for the zoo dataset.  The datasets are all available at the UCI repository \cite{Dua:2019}. Additionally, we performed experiments on a subset of the well-known MNIST dataset \cite{deng2012mnist} for digit classification. Here, we restricted the model to five classes (the digits  $0 \ldots 4$) and used 4500 (random) samples for training and 500 for testing. 

To model a classification task using a quantum circuit, first the data is encoded as a higher-dimensional binary vector.
Taking the iris dataset as a toy example, it consists of $4$ dimensional data encoding \textit{sepal length}, \textit{sepal width}, \textit{petal length} and \textit{petal width}.
After separating training and test data, a kMeans clustering on each dimension with $k=3$ is used on the training data. Thus, every datapoint can be encoded in a $4
\times 3=12-$dimensional binary vector which contains exactly $4$ non-zero entries. 
For the given cluster centers, the same can be done with the test data. Thus, a binary encoding is used to represent the datasets. Please note that binarized neural networks (BNNs), which are neural networks with weights and activations in $(-1,1)$, can achieve comparable test performance to standard neural networks, but allow for highly efficient implementations on resource limited systems \cite{NIPS2016_Hubara}.
In machine learning it is a common (and necessary) assumption that the training data belonging to one class shares commonalities, whereas it shares distinct properties to data from other classes. Clustering data has the goal to reduce the dimensionality of data points, while maintaining the internal data properties. Thus, naturally the points belonging to one class are clustered closer together while keeping them separated from data belonging to other classes. Our choice of preprocessing via classical clustering and binarization simplifies the classification task while encoding structure into the input. Our intention is to provide a uniform, hardware-friendly binary encoding across heterogeneous datasets and to stay within the constraints of current qubit counts and connectivity. 
 
Figure \ref{fig:ConvIris} shows the convergence behavior of the optimization algorithm during training the iris dataset. The loss decrease over iterations is shown in blue and the accuracy (of the test data) in red. In this run, the algorithm could perfectly classify all test samples.

\begin{table}
\centerline{
\begin{tabular}{|c|c|c|c|c|c|c|}
\hline
Dataset & Inp.& Perc. & qubits & Classes & Acc &Acc  \\
&&&&&(gd) & {\bf ours}\\
\hline
Iris & 12 & 4 & 19  & 3 & 84  $\%$ & 95  $\%$ \\
&&&&&($\pm$ 21) & ($\pm$ 5)\\ 
Zoo & 6 & 5& 18  & 7 & 72  $\%$& 87 $\%$ \\
&&&&&($\pm 22$)&  ($\pm 5$)  \\ 
wine & 12 & 3& 18  & 3 & 75  $\%$ & 95  $\%$ \\
&&&&&($\pm 5$)&  ($\pm 4$)  \\
mnist$_5$ &20 & 8 & 38 & 5& 75  $\%$ & 99  $\%$ \\
&&&&&($\pm 28$)&  ($\pm 5$)  \\
\hline
 \end{tabular}}
 
 \caption{Test datasets overview and performance. The column {\tt gd} indicates the performance when using a plain vanilla gradient descent approach for optimization. Please note the high variance of the results, due to local minima.}
 \label{tab:QML1}
 \end{table} 
 
Table \ref{tab:QML1} summarizes the four example datasets and the overall performance. It contains
the used qubits to represent the problem as quantum code, the amount of target classes and the achieved accuracy with gradient descent using autodiff ({\tt gd}) and our optimized quantum circuits ({\tt ours}).
 Please note the large variance of gradient descent, which results in a low average score due to local minima during optimization. Thus, when starting from a random seed, the optimizer often converges to good results comparable to our approach, but frequently the optimization fails completely. This effect of getting stuck in local minima is well known and documented in the machine learning literature \cite{picard2023torchmanualseed}. For this experiment, we did not perform data augmentation, dropout or other common methods to increase performance of gradient descent. 
 The combination of gradient descent with common measures to improve performance leads to similar results as our model.

\subsection{Hopfield Networks}
Introduced in 1982 \cite{Hopfield2554}, a Hopfield network consists of a one-layer recurrent network with the input dimension being equal to the output dimension. 
Its main purpose is to act as autoassociative memory,
e.g. for a given sample, the memory should retrieve a piece of memorized data.  Hopfield networks can be applied in de-noising or removing interference from an input or they can be used to determine whether the given input is \textit{known} or \textit{unknown}.
Although Hopfield networks are a well-established topology, they
have received renewed attention recently \cite{ramsauer2021hopfield}. 
\begin{figure}
  \includegraphics[width=0.45\textwidth]{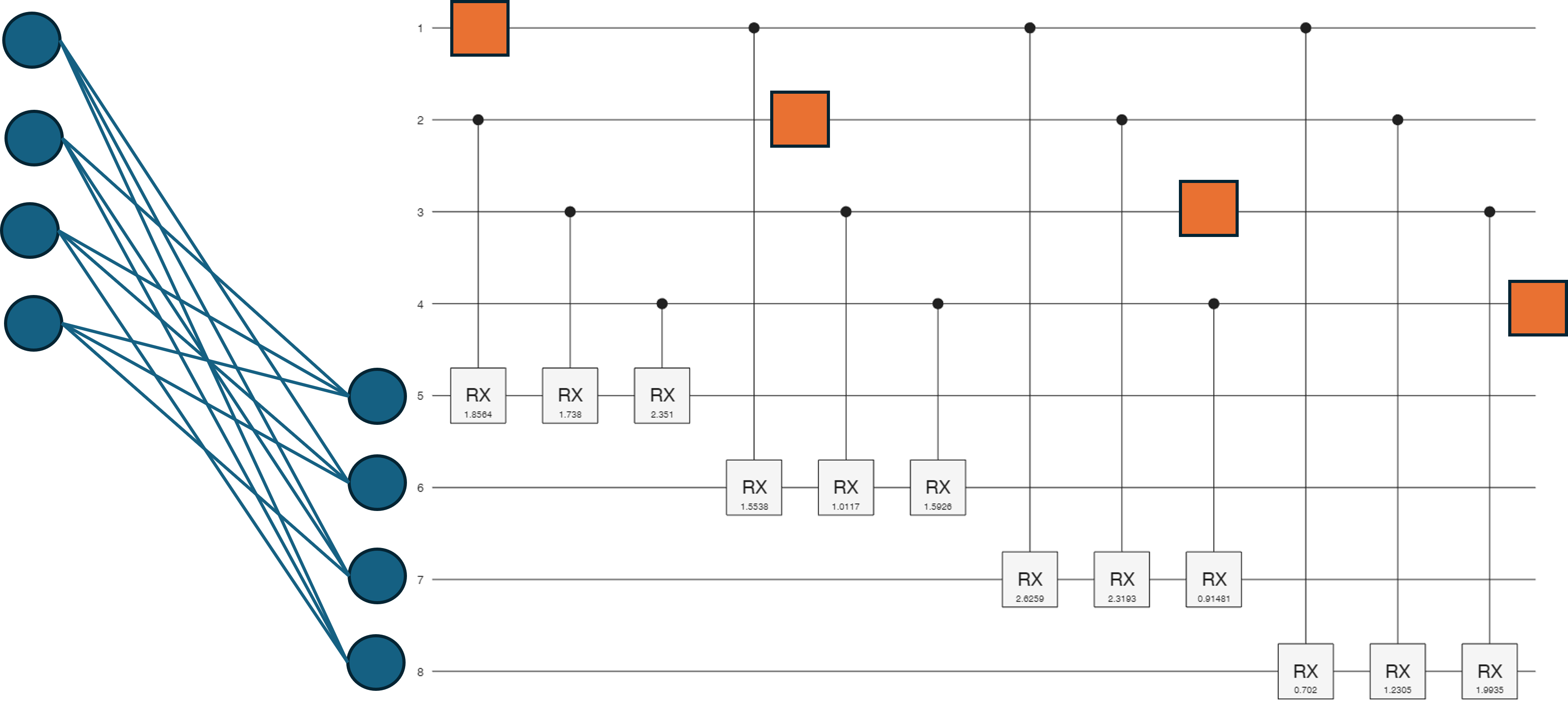}    
\caption{Hopfield Network and its corresponding Quantum Circuit for four input qubits. The red boxes are visual marks to show that an input artificial neuron has no direct connection to its corresponding output neuron.
}
\label{fig:HopfieldNN}
\end{figure} 
\begin{figure}
  \includegraphics[width=0.45\textwidth]{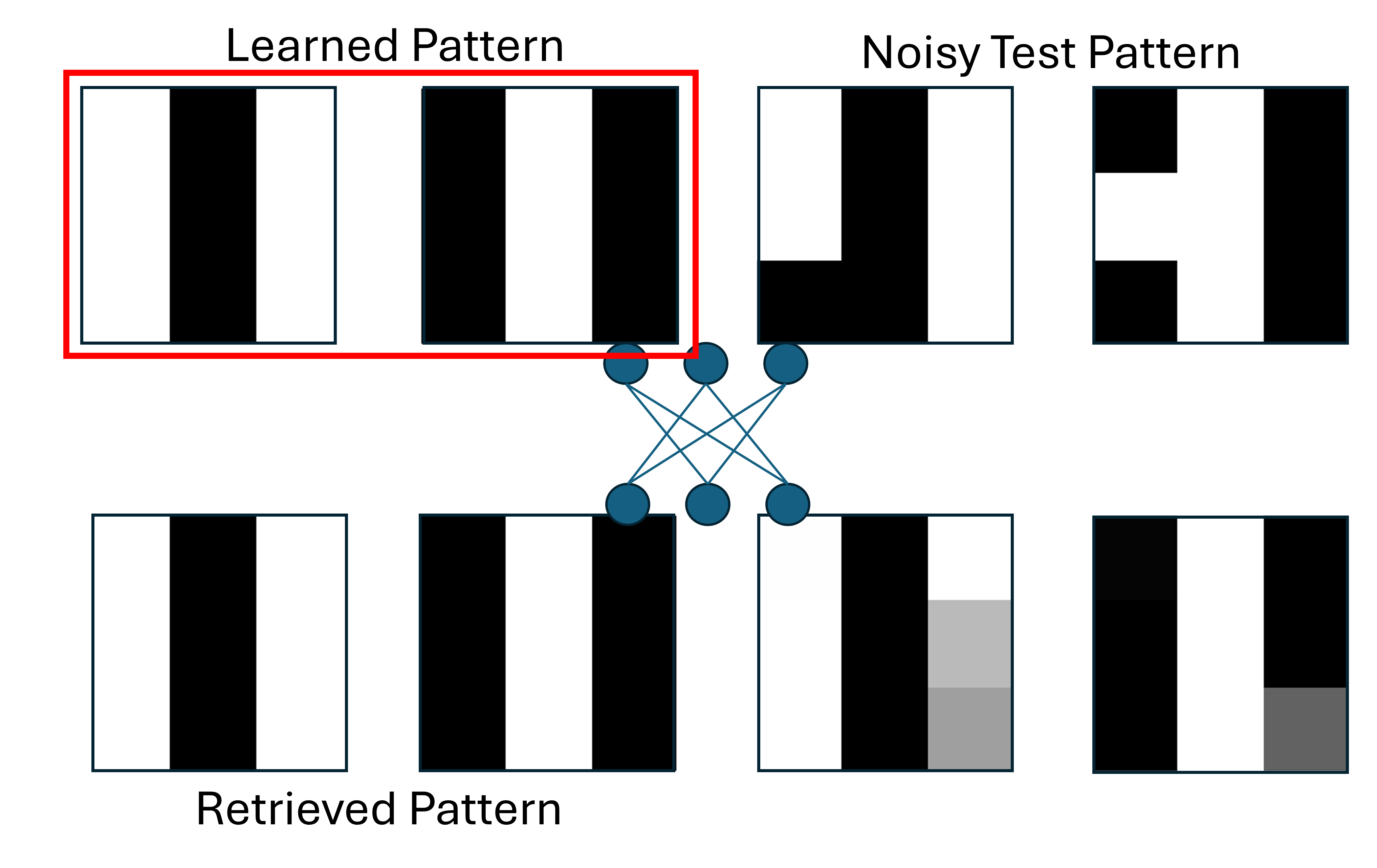}    
\caption{Pattern retrieval of a 9-dimensional input data. The patterns in the upper left (red box) are to be memorized from the Hopfield network. The first row shows the test pattern and the second row the retrieval pattern (after one iteration). Whereas the memorized patterns are successfully reconstructed, the noisy input patterns have activation probabilities leading to the closest memorized pattern. 
}
\label{fig:HopfieldExp}
\end{figure} 
 \begin{figure}
  \includegraphics[width=0.45\textwidth]{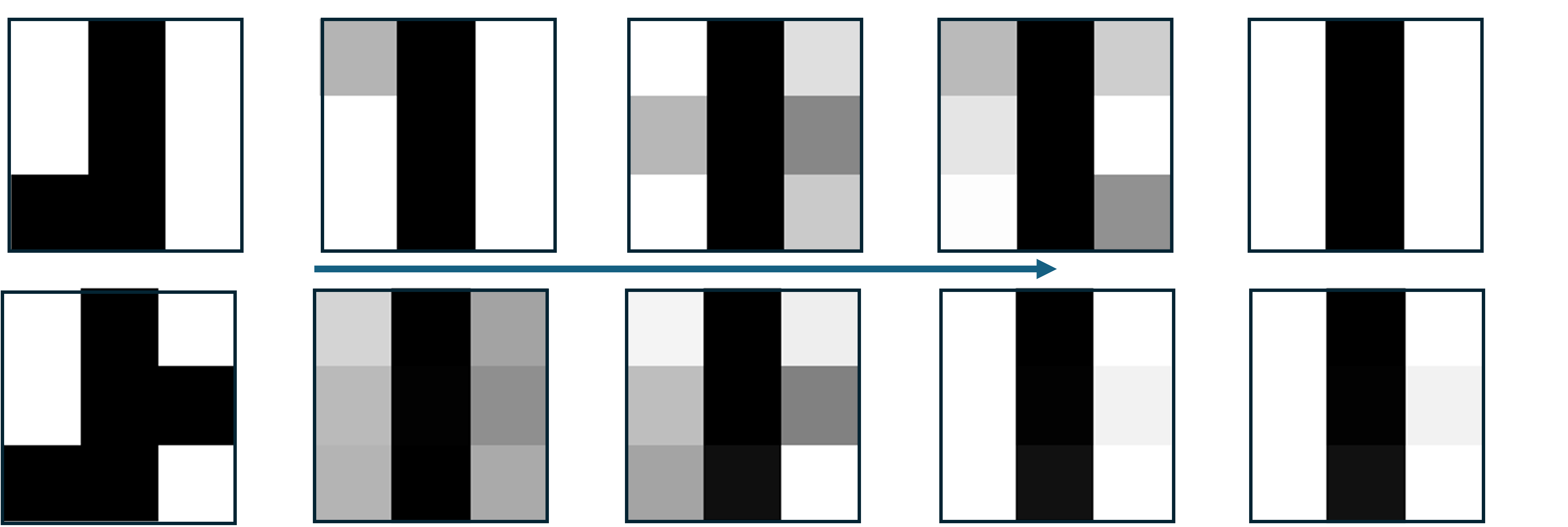}    
\caption{Pattern retrieval of disturbed input data (left). As recurrent model, the optimized Hopfield network can recover the memorized pattern within a few model iterations (right).
}
\label{fig:HFIter}
\end{figure} 
The units in Hopfield nets are binary threshold units $\in \{0 ,1  \}$,
the interactions $\omega_{i,j}$ between neurons
are defined to be symmetric 
($\omega_{i,j}=\omega_{j,i}$) and no unit has a connection with itself ($\omega_{i,i}=0$).
The classical form of a Hopfield network is shown on the left of Figure \ref{fig:HopfieldNN}. The corresponding expression as quantum code is depicted on the right; the red squares mark that there is no connection from an input dimension to its output dimension. Thus, the remaining artificial neurons have to determine the activation of a certain input value. Hopfield networks are recurrent, which means that the output is fed back as input and the process is iterated until convergence to a memorized pattern.
Figure \ref{fig:HopfieldExp} summarizes an experiment for memorizing and retrieving patterns. The goal is to learn two
patterns (of vertical stripes) of a 9-dimensional ($3\times 3$) input data. They are depicted in the red box at the upper left. After optimization of the model we can use it for inference.
The first row in Figure \ref{fig:HopfieldExp} shows the test pattern and the second row the retrieval pattern (after one iteration). Whereas the memorized patterns are successfully reconstructed, the noisy input patterns have activation probabilities leading to the closest memorized pattern. 

Figure \ref{fig:HFIter} shows the evolution towards a memorized pattern over several iterations of the recurrent model during the quantum simulations. In our experiments the model typically converges after three to five iterations.

\subsection{Restricted Boltzmann Machines}
\begin{figure}
  \includegraphics[width=0.45\textwidth]{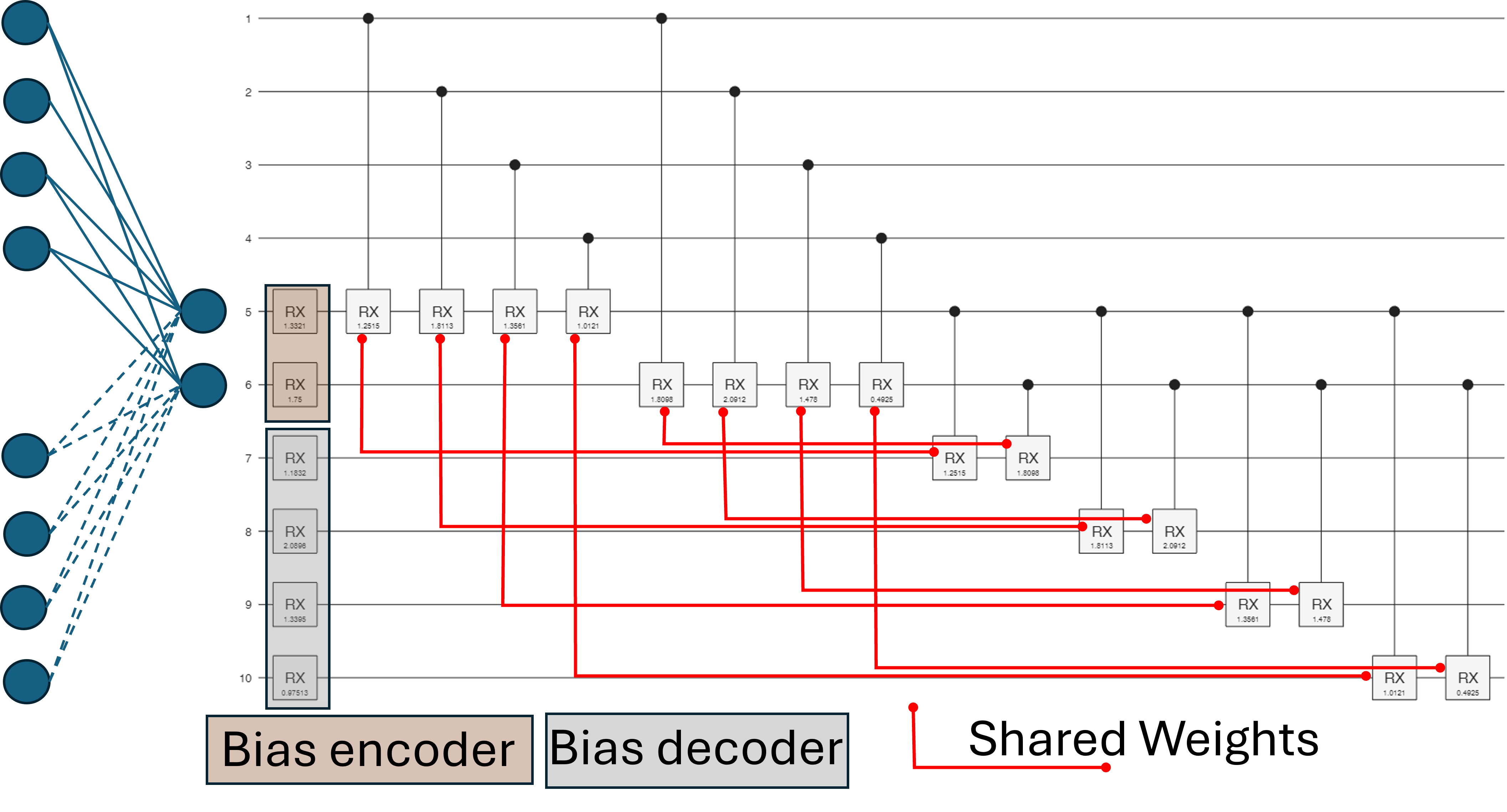}    
\caption{RBM Network and its corresponding Quantum Circuit for four input qubits and two latent space artificial neurons. The red connections indicate a weight sharing. The RBM is bidirectional, thus the weights are used for both, encoding and decoding.
}
\label{fig:RBM}
\end{figure} 
Some decades after the introduction of the perceptron, around 1985, the Boltzmann Machine (BM) was invented
\cite{Ackley1985ALA}. It is a network of symmetrically connected, neuron-like binary units. Learning the weights
of such a connectionist system allows the Boltzmann Machine
to discover features that represent complex properties in training data.
A modification is a so-called
Restricted Boltzmann Machine (RBM) which consists of a two layer architecture with one
layer of visible units and one layer of hidden units. 
RBMs were initially invented under the name \textit{Harmonium} by Paul Smolensky in 1986 \cite{Smolensky86}.
An RBM can be interpreted as a bipartite graph with symmetric connections, see Figure \ref{fig:RBM}. 
\begin{figure}
  \includegraphics[width=0.45\textwidth]{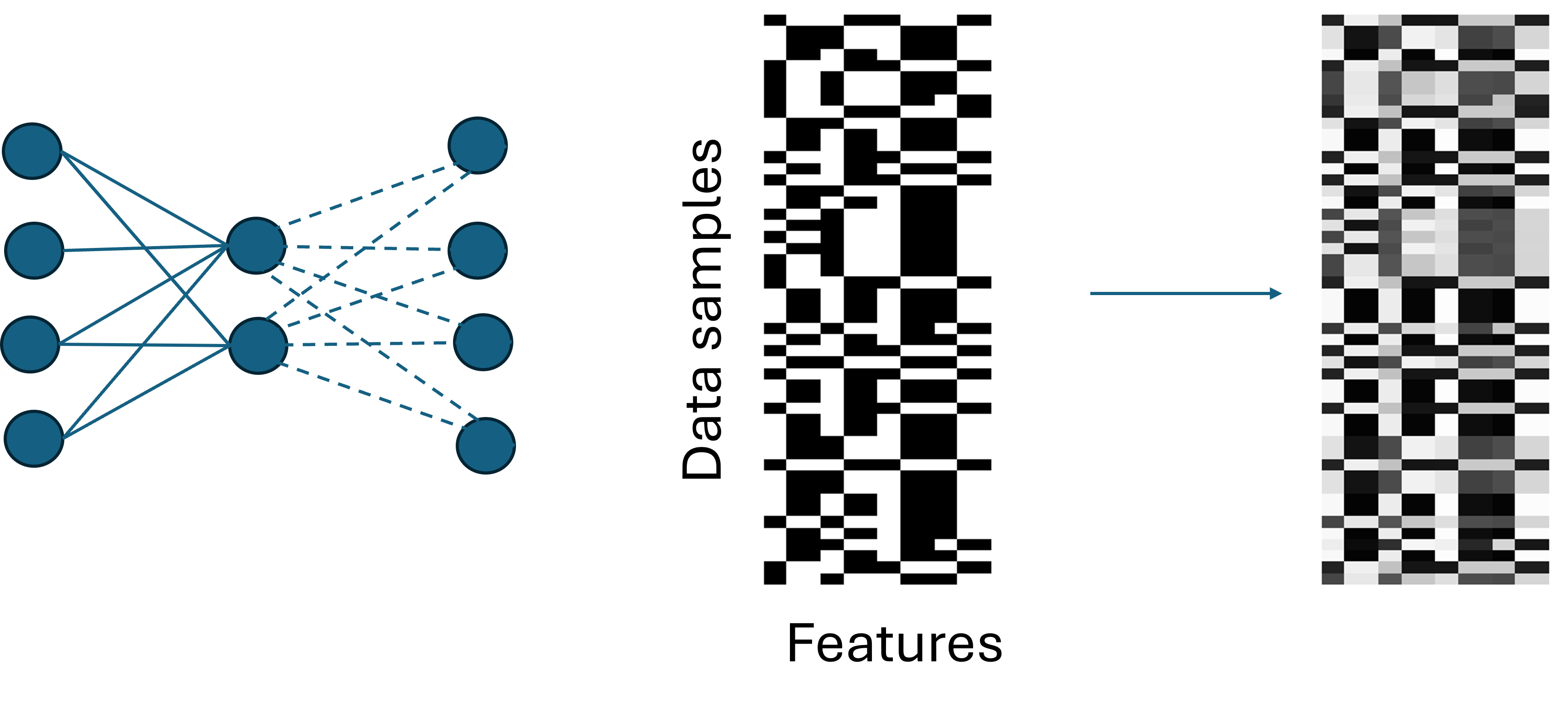}   
\caption{Example features from the Iris dataset and its encoding and decoding. Since the RBM is a probabilistic model, the gray values indicate the probability for a perceptron activation.  Thus, already two hidden perceptrons are sufficient to represent the 12-dimensional input space with reasonable quality.
}
\label{fig:RBMRec}
\end{figure} 

Such a model can be useful for dimensionality reduction, classification, regression, collaborative filtering, feature learning and topic modeling.
Indeed, Restricted Boltzmann Machines received increasing attention during and after the \$1 million Netflix challenge
\cite{Salakhutdinov07}.
In contrast to deterministic models, RBMs are generative stochastic networks, where the artificial neuron activation is probabilistic. Thus, RBMs have binary units which turn \textit{on/off} according to probabilities determined by the weights, which is a model fitting perfectly to our quantum formulation.
The units in Restricted Boltzmann Machines are binary threshold units $ \in [0 ,1 ] $,
the interactions $\omega_{i,j}$ between artificial neurons
are defined to be symmetric. The weights need to be suited for the projection onto the hidden space and vice versa the reconstruction from the hidden space to the input space. Thus, for an unsupervised learning task, the RBM can be unrolled to an autoencoder like architecture with shared weights among the encoder and decoder, indicated by the dashed connections in Figure \ref{fig:RBM}. This has also been pointed out by \cite{Smolensky86}. Similar to the Hopfield architecture, the shared weights decrease the complexity of the stochastic optimization procedure we described beforehand, as the amount of variables is effectively decreasing. As a simple example, we use the iris dataset with a binary representation of dimension 12, so 12 qubits are used as input values. We then optimize for a two dimensional latent space and uplift the representation to 12 dimensions again, as visualized in Figure \ref{fig:RBM} (just using 12 instead of four input qubits shown in the image). Figure \ref{fig:RBMRec} shows on the right the reconstruction quality of (unseen) data samples after optimizing the model. As can be seen, even though there is a heavy compression enforced (to two dimensions), the reconstruction is of reasonable quality. 
 
\subsection{Autoencoder}
\begin{figure}
  \includegraphics[width=0.45\textwidth]{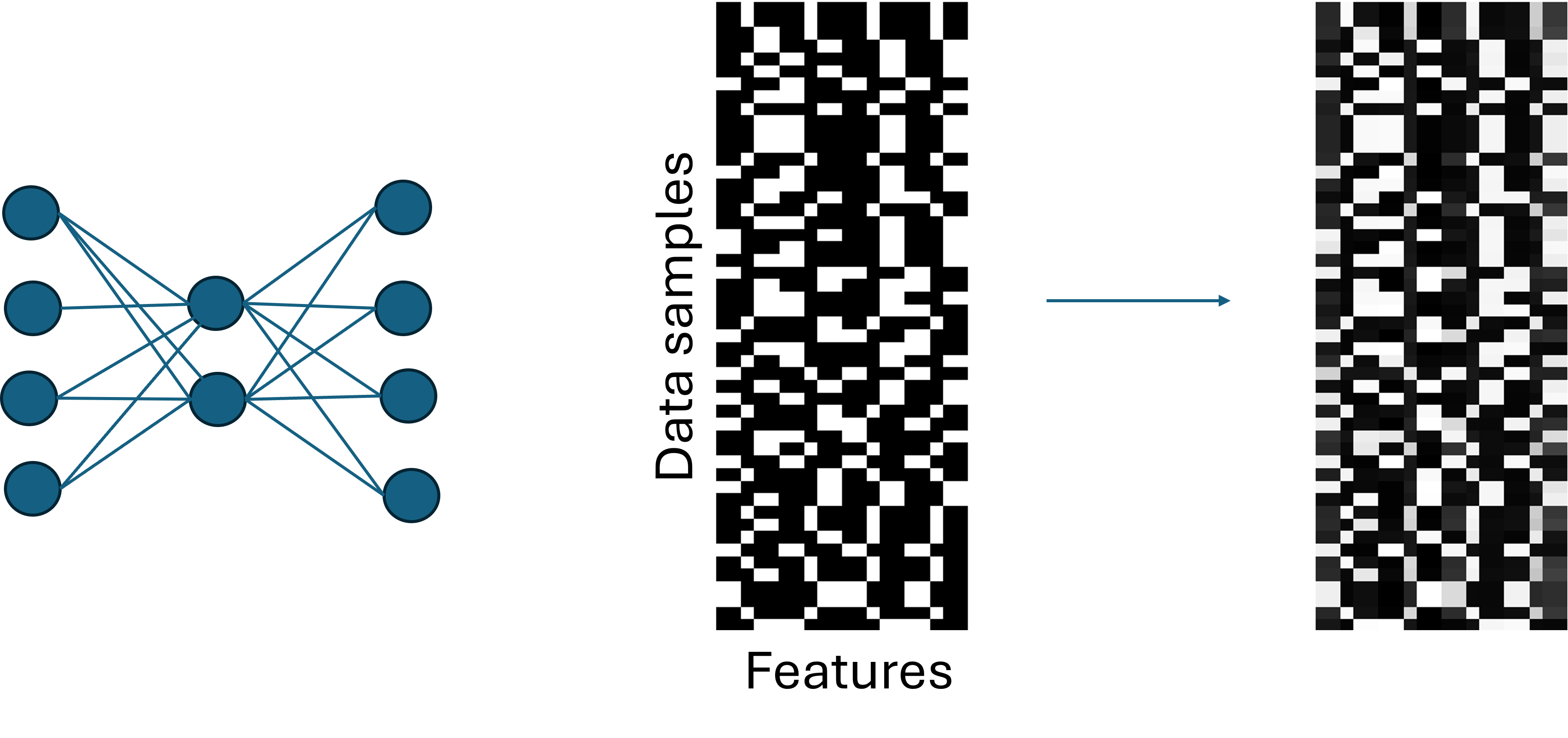} 
\caption{Example features from the Iris dataset and its encoding and decoding using an autoencoder. Since the autoencoder can carry more information than an RBM, the overall reconstruction quality is higher, compared to the former model. 
}
\label{fig:AE}
\end{figure}

An autoencoder is a neural network which is commonly used to learn efficient representations from data \cite{doi:HintonAE}. It consists of two parts, an encoder branch and a decoder branch with a bottleneck layer in the middle. The aim of the model is to solve a copying task by replicating the input signal at the output. Due to the bottleneck layer, which has usually less dimensions than the input data, the model has to learn a compressed representation of the training data \cite{Bank2023}. Autoencoders have been modified to variational autoencoders 
\cite{kingma2022autoencodingvariationalbayes}, sparse autoencoders \cite{huben2024sparse}, vector quantized autoencoders \cite{VQVAE19} and more in the past. Still, an autoencoder is a common approach for data compression, subspace projection and data analysis. 
The implementation of a quantum circuit to realize an autoencoder is very similar to the circuit shown in Figure \ref{fig:RBM}. The only difference is that weight sharing is removed, which means that there are more parameters to optimize (which in return can take longer during training), but the separated weights for encoding and decoding allow for a higher capacity for representing data. We performed a similar experiment on the Iris dataset used in the RBM framework for the autoencoder. Again, a 12 dimensional input vector is reduced to two dimensions and reconstructed back to 12 dimensions. The outcome of the reconstruction  is shown in Figure \ref{fig:AE}. It can be seen that the reconstruction quality improved over an RBM, shown in Figure \ref{fig:RBMRec}. This expected result comes from the dropped weight sharing which allows the model to separate information for encoding and decoding. This leads to a higher capacity of the neural network.

\subsubsection{RBM and AE Model Capacities}

\begin{figure}
  \includegraphics[width=0.45\textwidth]{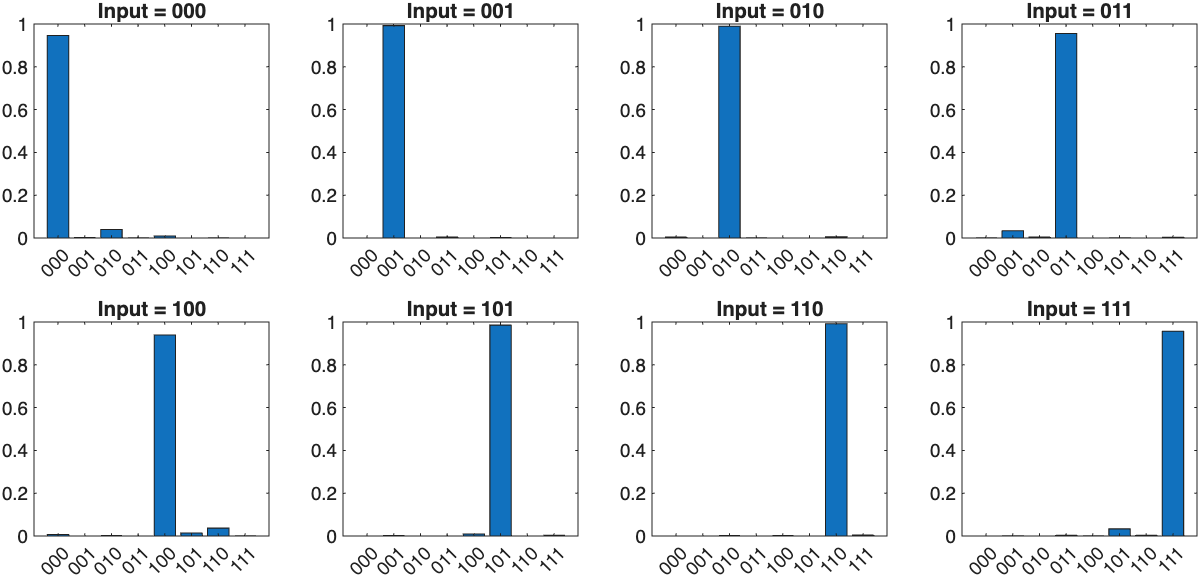}  
\caption{Input/Output to learn all 3-dimensional binary patterns. The subtitle shows the input data, which can be observed with a high likelihood when executing the quantum circuit. To represent this dataset, a three dimensional latent space is at least required. 
}
\label{fig:AEInOut}
\end{figure}

In the following experiment, the aim is to compare the capacities and convergence behavior of a Restricted Boltzmann Machine (RBM) and an Autoencoder (AE) for varying latent space dimensions. In our first experiment, the data to learn is the $2^3$ dimensional binary code. This means for an input $000$ the RBM and the AE should return $000$ while passing the input data through the bottle neck layer for dimensions one to four. It is expected that for an autoencoder, a one or two dimensional latent space will not be able to recover the data, whereas three or four dimensional latent space is sufficient. 
Since the dataset allows strict bipartite energy interaction, both models should be able to reproduce the dataset with 3 and more hidden neurons.
  Figure \ref{fig:AEInOut} shows the observation likelihoods for a trained model with sufficient capacity. 
The subtitle shows the input code and the histogram shows the observation likelihood of a model when executing the quantum circuit. To represent this dataset, a three dimensional latent space is at least required.

\begin{figure}
  \includegraphics[width=0.45\textwidth]{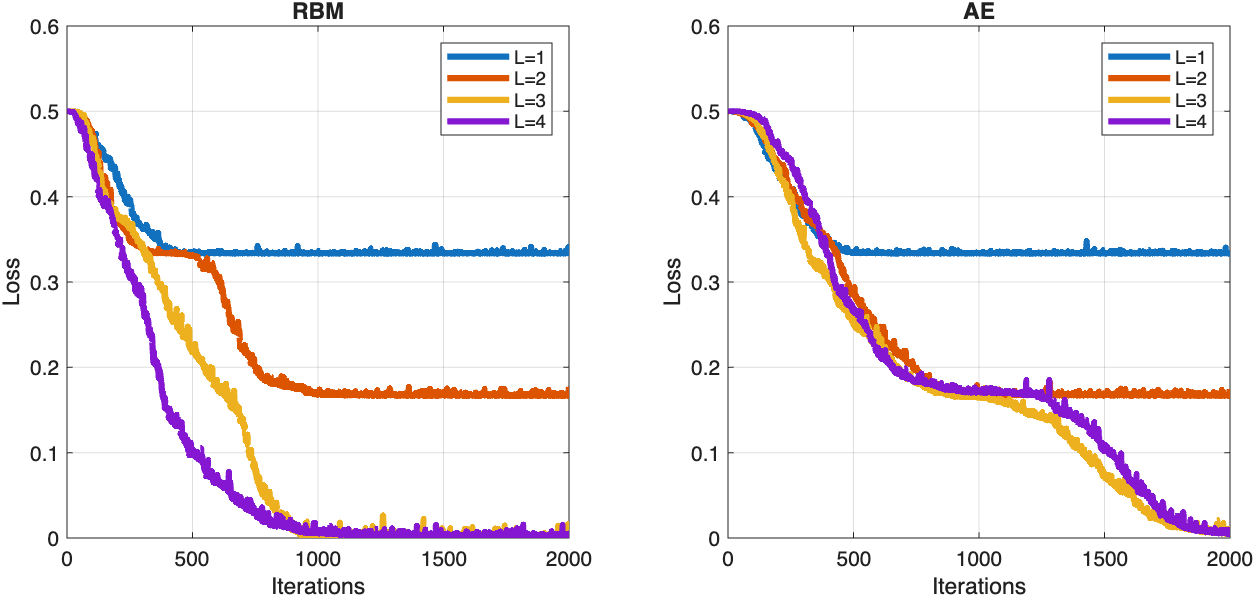}  
\caption{Training data: All $2^3$ 3-dimensional binary digits from $000$ to $111$. Left: Optimization of a RBM with a latent space of 1-4 dimensions. Right: Optimization of an autoencoder with a latent space of 1-4 dimensions. Both models can capture the data with 3 and more latent space dimensions.
}
\label{fig:RBMAEIter}
\end{figure} 

\begin{figure}
  \includegraphics[width=0.45\textwidth]{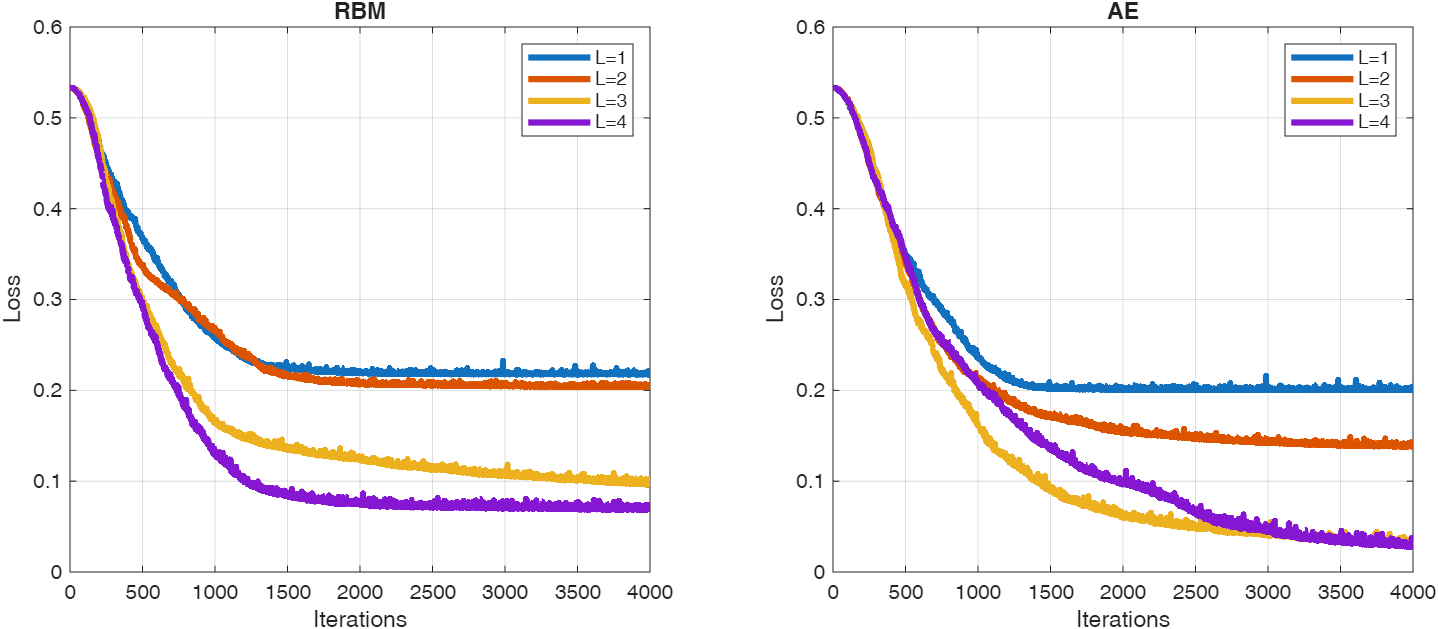}  
\caption{The training data (see text) consists of five 6-dimensional (asymmetric) patterns.  Left: Optimization of a RBM with a latent space of 1-4 dimensions. Right: Optimization of an autoencoder with a latent space of 1-4 dimensions. The asymmetry of the training data is very hard to capture for a RBM, whereas an AE can represent the data. This example shows that an AE has in general a higher capacity over a RBM.
}
\label{fig:RBMAEIter2}
\end{figure} 

Figure \ref{fig:RBMAEIter} summarizes the optimization of the dataset for both, a RBM and an AE for different latent space dimensions. 
The x-axis shows the iterations during optimization and the y-axis the loss. The loss is defined as the reconstruction quality. To determine this, we simulate the quantum circuit. For a target $T$ we compute 
$\sum_i(prob(i)*xor(state(i),T)$, with
$xor$ denoting the Hadamard distance between the observed variable and the target $T$, multiplied with the observation likelihood. 
Thus, if the state(i) is equal to $T$ and observed with a probability $1$, the score is zero. This score is evaluated and summed up for all training data (all binary codes). If all states can be reconstructed perfectly by the RBM or AE, the overall loss is zero.
All involved weights are initialized with zero. Thus, the models start their optimization with the same loss.

The left diagram shows the loss of the RBM, whereas the right side shows the AE. It can be seen that both models require a latent space of three to reconstruct the data perfectly, as three perceptrons can be used to represent the $2^3$ dimensional binary space perfectly. When the latent space is higher, the amount of variables increases, which means that the convergence is slower. 
Since in a RBM less variables are optimized, the convergence is faster and requires less iterations, compared to a similarly structured autoencoder. 

Figure \ref{fig:RBMAEIter2} summarizes the optimization of a very simple dataset $D$ with 5 examples and 6 dimensions, namely:
\begin{eqnarray}
D&=&\begin{pmatrix}
1 & 1 & 1 & 0 & 0 & 0 \\
0 & 0 & 0 & 1 & 1 & 1 \\
1 & 0 & 1 & 0 & 1 & 0 \\
0 & 1 & 0 & 1 & 0 & 1 \\
1 & 1 & 0 & 1 & 1 & 0
\end{pmatrix}
\end{eqnarray}
This dataset has the property that it has asymmetries and therefore it is expected that the RBM will perform worse than the autoencoder. 
Similar to Figure \ref{fig:RBMAEIter}, the x-axis shows the iterations during optimization and the y-axis the loss. The left hand side shows the RBM for different latent space dimensional and the right hand side the autoencoder.
As the dataset contains more asymmetries it demonstrates that the AE has a higher capacity than a RBM. This comes from the strict bipartite energy interaction the RBM requires. Thus, the possible expressiveness is higher for an autoencoder. All these observations are expected and demonstrate that our quantum neural network behaves as expected. 

\subsection{Convolutional Neural Network}
\begin{figure}[b]
  \includegraphics[width=0.45\textwidth]{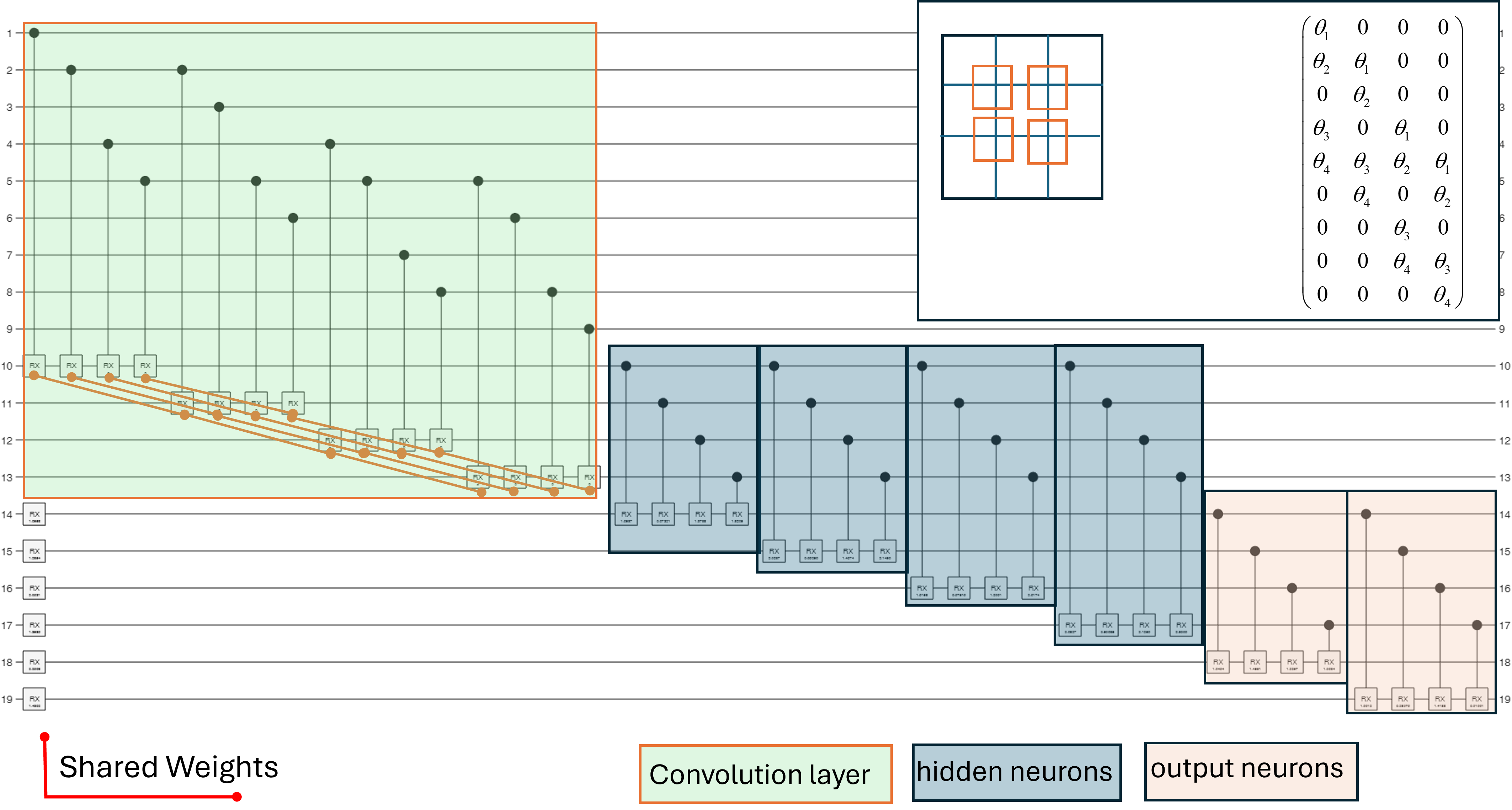}   
\caption{Convolutional neural network 
}
\label{fig:ConvTopo}
\end{figure} 

A convolutional neural network (CNN) is a feedforward neural network that learns features by optimizing for linear shift-invariant filters \cite{LeCun2015}. A core concept is a so-called weight sharing and connection cutting which leads to a linear form of a Toeplitz matrix to be optimized \cite{bookboettcher}. 
\begin{figure}
  \includegraphics[width=0.35\textwidth]{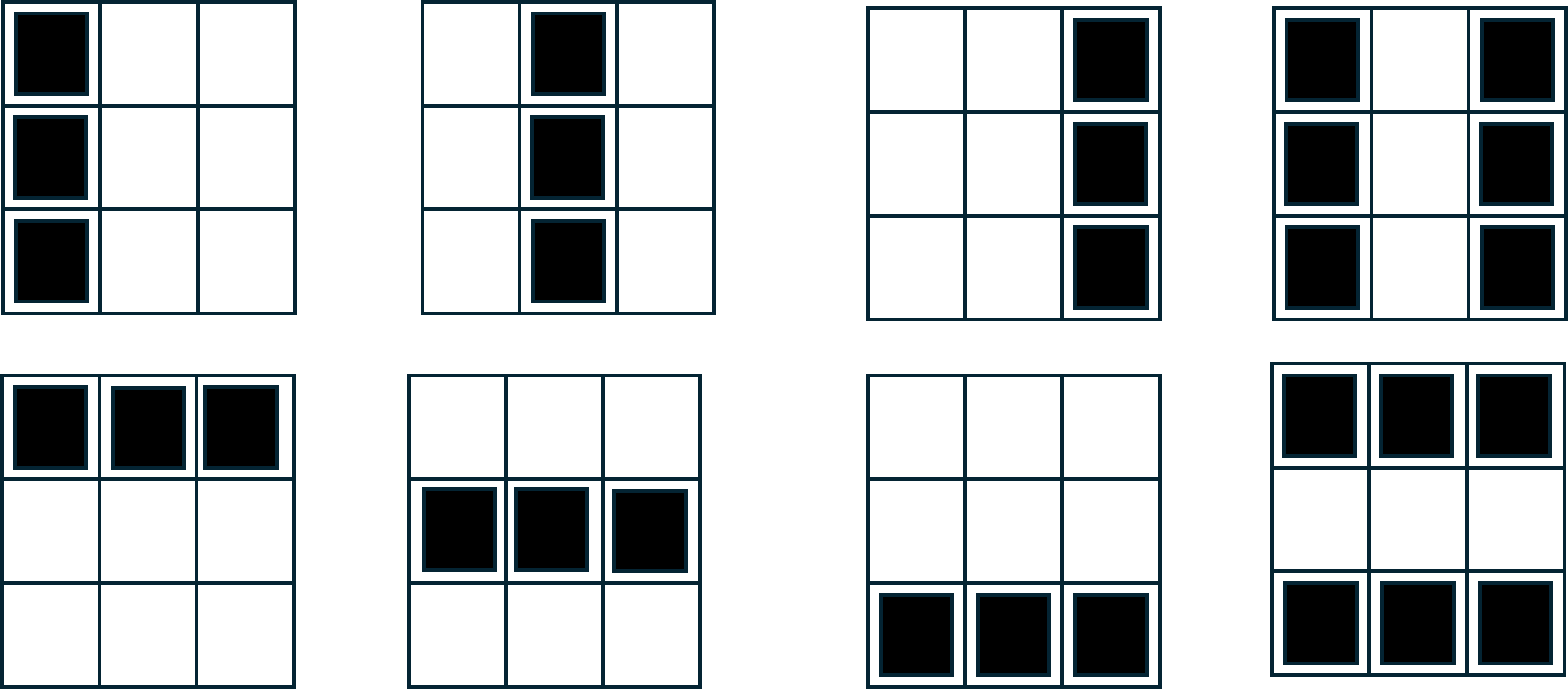}   
\caption{The Bars and Stripes patterns.
}
\label{fig:BarsStripes}
\end{figure} 
This type of deep learning network has been applied to many different types of data including text, images and audio \cite{9758454,9451544}. It can be seen as the
 de facto standard in deep learning-based computer vision and has only recently been replaced by alternative architectures such as the transformer \cite{LIN2022111}. 
 The concept of a convolution is visualized in the upper right corner in Figure \ref{fig:ConvTopo}. A $2 \times 2$ kernel is placed over an image patch, the image values are multiplied with the kernel weights and summed up. The kernel is then shifted to the next patch and processed with the same weights. Thus, it is a linear shift-invariant operation which can be expressed as a fully connected layer with several connections being dropped and the weights being arranged as a Toeplitz matrix. This again leads to a high reduction of optimization variables which makes the optimization simpler, compared to several fully connected layers. Thus, overfitting is more easily prevented and it is possible to learn concepts for classification in an easier way.
 Figure \ref{fig:ConvTopo} shows the quantum code of a convolutional neural network with one convolution layer in the beginning and then a fully connected layer before continuing to the output layer. We used this model to train simple pattern classification tasks, such as bars and stripes, see Figure \ref{fig:BarsStripes}: The input is a $3 \times3$ patch and the task is to classify if the input pattern belongs to the bars and stripes examples or not. After optimizing the model, the network can perfectly classify and differentiate the examples. Please note that the task is quite imbalanced, as there are only $8$ positive patterns and $2^9-8$ negative patterns. Such imbalanced data can pose major challenges in machine learning. Still, by using the Kiefer–Wolfowitz algorithm and preferencial sampling \cite{j.1467-9876.2009.00701.x} we can obtain a classifier which classifies all patterns correctly, when using majority voting.

\subsection{Generative Neural Network}
\begin{figure}
  \includegraphics[width=0.45\textwidth]{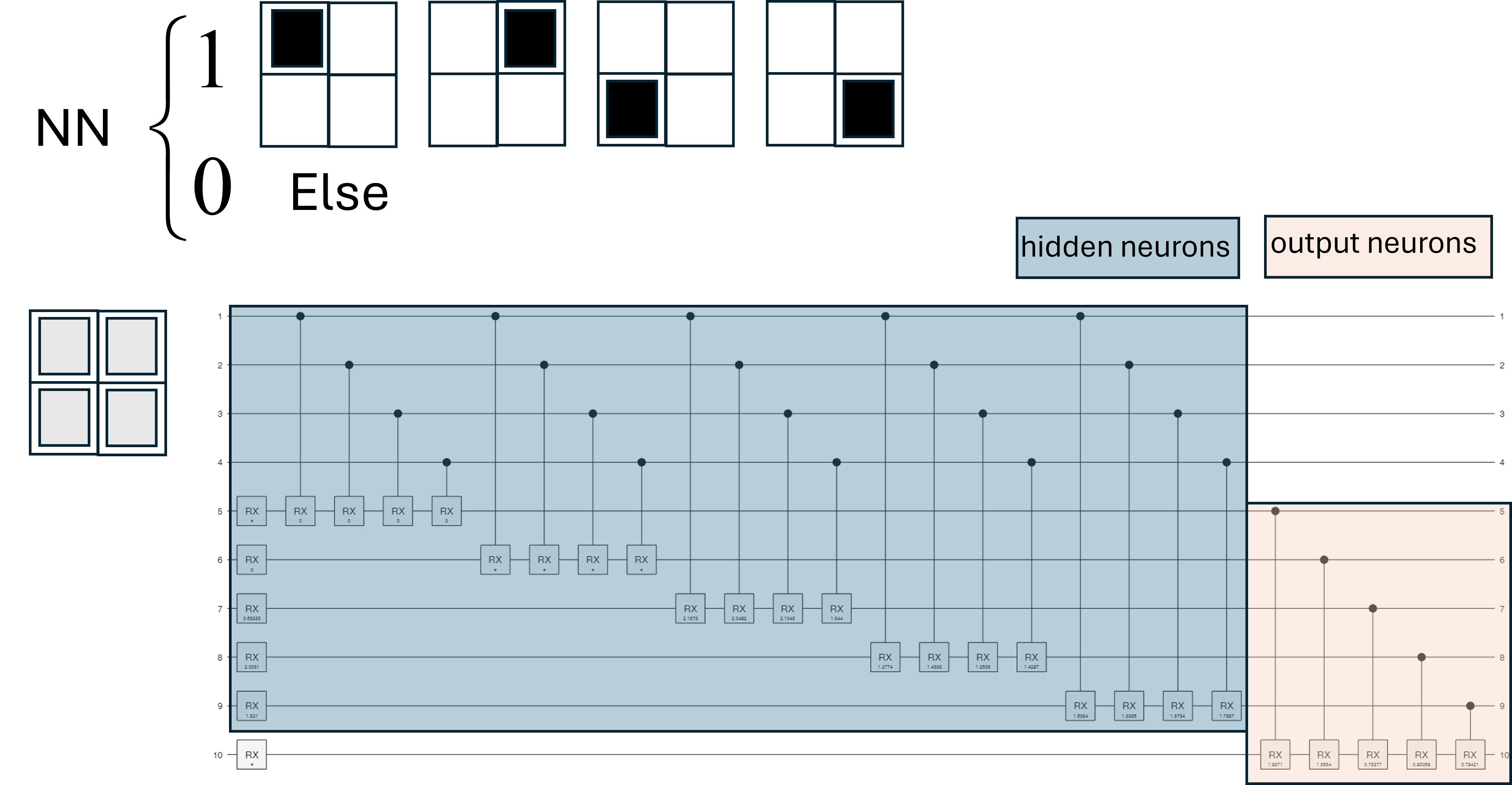}   
\caption{Binary neural network for classifying a simple pattern 
}
\label{fig:BinTopo}
\end{figure} 
\begin{figure}
  \includegraphics[width=0.45\textwidth]{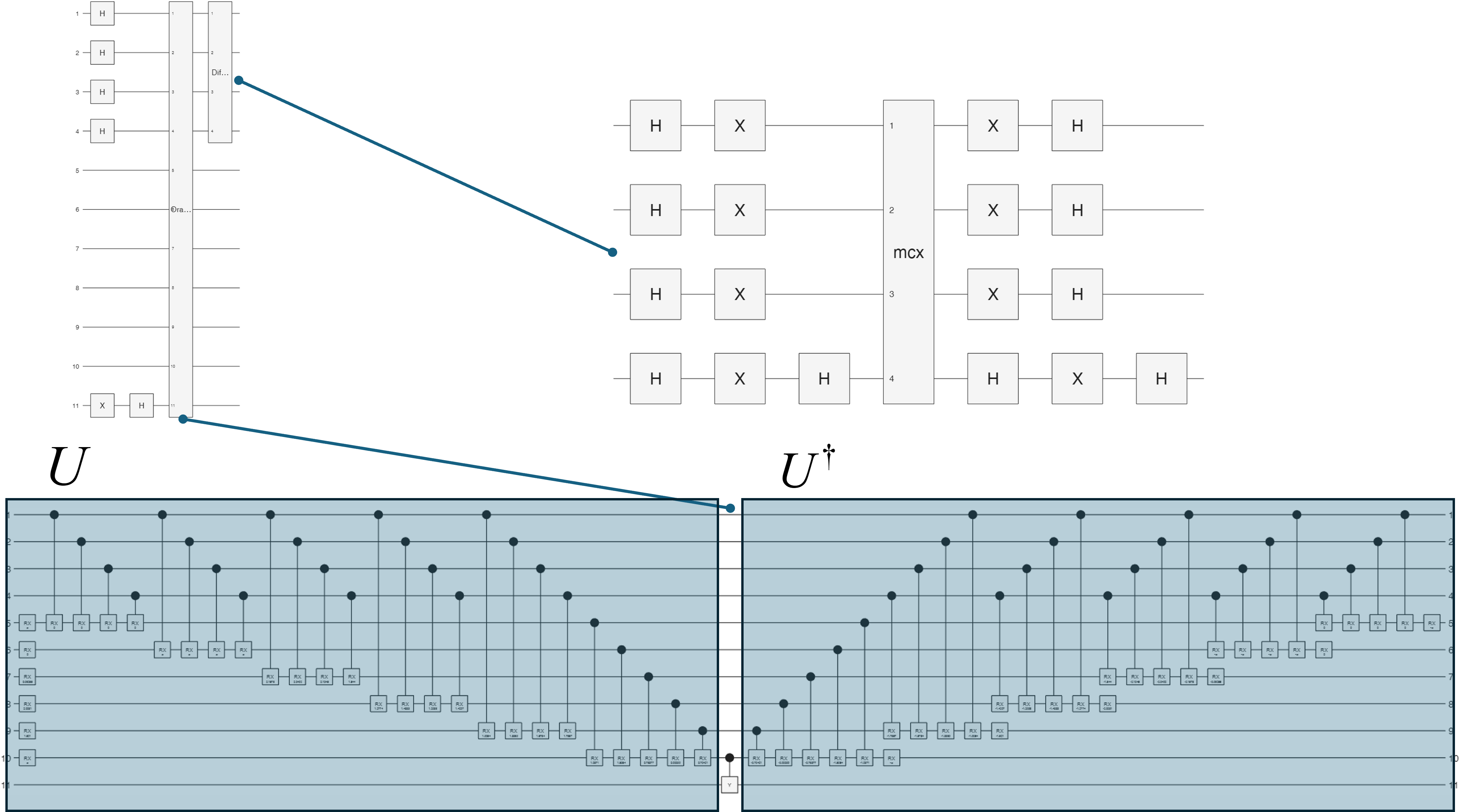}   
\caption{Architecture of the Grover circuit for generating patterns.
}
\label{fig:GroverTopt}
\end{figure} 

\begin{figure}
  \includegraphics[width=0.45\textwidth]{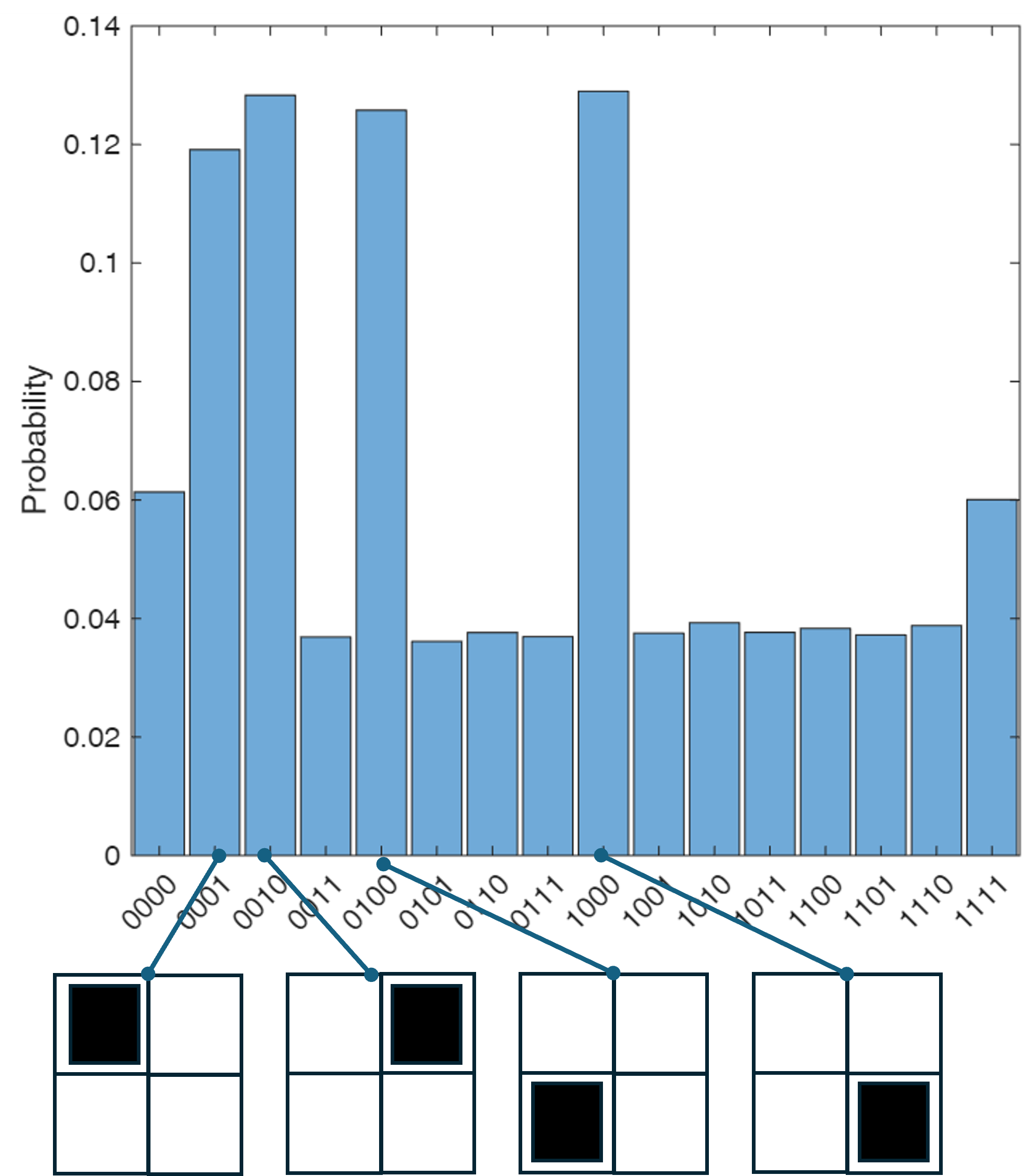}   
\caption{Sampled bins when executing Grovers algorithm
}
\label{fig:GroverBin}
\end{figure} 
\begin{figure}
  \includegraphics[width=0.45\textwidth]{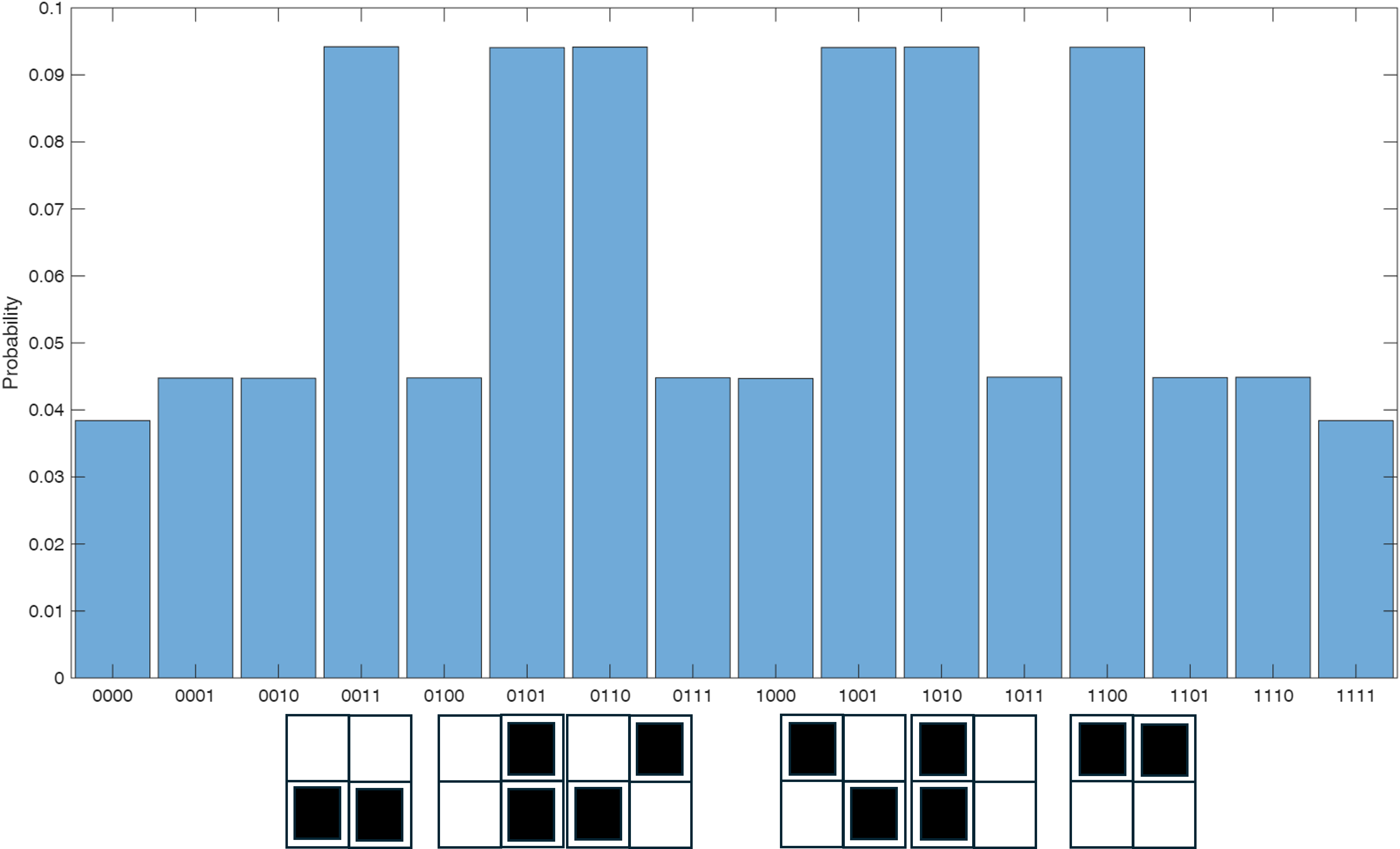}   
\caption{Histogram after executing Grovers algorithm for a neural network which detects patterns with two dots.
}
\label{fig:GroverBin2}
\end{figure} 
Generative AI (GenAI) is an increasingly important field of research and comprises the challenge to use data to generate new results or contents such as texts, speech, audio recordings, images or videos. Early works date back to 2002 \cite{NIPS2001_7b7a53e2,Blei03} with the real breakthrough in using generative adversarial networks \cite{NIPS2014_f033ed80,NEURIPS2018_8cea559c}, also called GANs, vector quantized variational autoencoders \cite{Oord17}, denoted as VQ-VAEs and the so-called diffusion models \cite{DiffSurvey23, rombach2022high}. The core idea of a GAN is to train two neural networks which compete in a min-max game, where one model, the so-called \textit{generator}, tries to fool the other model, called the \textit{discriminator}. Whereas the discriminator has to distinguish real data from generated data, the  generator takes a random pattern as input to generate an artificial sample. Typical challenges for GANs are the unstable training behavior and mode collapse. This led to alternative approaches, such as diffusion models.  A diffusion model has two main parts, first the forward diffusion process, and secondly the reverse sampling process. The goal of a diffusion model is to learn a diffusion process such that generated elements are distributed in accordance with an original dataset. In practice, e.g.\ for image generation, the diffusion model is trained in such a way to reverse the process of adding noise to an image. Thus, 
when starting with an image composed of random noise, the network is applied iteratively to denoise the image, leading to novel examples. We do not go into details on the vast amount of literature, instead we refer to recent surveys, such as \cite{gozalobrizuela2023surveygenerativeaiapplications,Schneider2024,Sengar2025}.
All approaches are characterized by involving a stochastic process, either for a denoising of a random pattern (in diffusion frameworks), or for a min-max game to fool a discriminator, e.g. when using a GAN. 

In the last years, the Grover algorithm \cite{Grover01} has gained increased attention in quantum computing as it is
 a celebrated quantum algorithm that provides a provable quadratic speedup over classical algorithms for searching in an unsorted database. 
Especially for quantum annealing \cite{LucasAnneal14,APOLLONI1989233} and quantum machine learning \cite{10.1007/s11227-022-04923-4,AbuGhanem2025,9798970}, the Grover algorithm is a very common tool to use.

The core idea for a quantum GenAI circuit is to use our quantum neural networks (e.g. for classification of patterns) to combine it with a Grover search, such that examples can be sampled which fulfill a classification property. Figure \ref{fig:BinTopo} shows the architecture of our quantum neural network to classify a simple binary pattern. Here the goal is to classify, if there is exactly one dot in the pattern or not. After training the model we keep the circuit frozen and generate the circuit and its reverse to get the so-called oracle, see Figure \ref{fig:GroverTopt}. 
Now we generate a circuit by first bringing the input qubits into superposition. Then the oracle is applied and afterwards a diffusion circuit follows. Now, we can execute the circuit and visualize the probabilities for sampling specific patterns. A resulting example distribution is shown in Figure \ref{fig:GroverBin}. Thus, as expected, the classified patterns are sampled with a much higher likelihood than the remaining patterns. Figure \ref{fig:GroverBin2} shows the result for a neural network which can detect two bit patterns.
Instead of a simple $2 \times 2$ pattern, imagine a frozen quantum network which can classify whether an image depicts a human face or not. In combination with superposition in front and the Grover circuit afterwards, the sampled  proposals are more likely to be classified as a face. Thus, sampling this circuit leads to a highly-efficient generative model.

\section{Conclusion}
In this work we presented a formulation to express and optimize stochastic neural networks as quantum circuits for gate-based quantum computing. Our work is inspired by a classical perceptron and we use CRX-Gates to drive the stochastic activation of an  artificial neuron, but it is easily possible to extend the perceptron with CRY- and CRZ-Gates in case of complex signals. A combination of several such stochastically activating neurons is used to express a quantum neural network. The Kiefer-Wolfowitz algorithm in combination with simulated annealing is used for training the network weights. It has the advantage that it can easily respect constraints on the weights, such as shared weights, connection cutting and more. This allows in a very intuitive way to optimize
several topologies and models. Our experiments present shallow fully connected networks, Hopfield Networks, Restricted Boltzmann Machines, Autoencoders and convolutional neural networks. We finally demonstrate the combination of our optimized neural networks as an oracle within the Grover algorithm to present a circuit that allows for a quantum-driven generative AI.  
\subsection*{Funding Declaration}
This work was supported, in part, by the Federal Ministry of Research, Technology and Space (BMFTR), Germany under the AI service center KISSKI (grant no. 01IS22093C), the QC service center QUICS (grant no. 13N17418),
by the Quantum Valley Lower Saxony and by Germany's Excellence Strategies EXC-2122 PhoenixD and EXC-2123 Quantum Frontiers.
\bibliographystyle{plain}
\bibliography{sample}

\end{document}